\documentclass[prd, 10pt,aps, floatfix, notitlepage,twocolumn,longbibliography,nofootinbib]{revtex4-1}

\usepackage{amsmath,amsfonts,amssymb}
\usepackage{comment}
\usepackage{mathrsfs}
\usepackage{graphicx}
\usepackage[english]{babel} 
\usepackage{url} 
\usepackage{xcolor}
\usepackage{bbold}
\usepackage{adjustbox}
\usepackage[utf8]{inputenc} 
\usepackage[colorlinks=true,citecolor=blue,urlcolor=blue]{hyperref}
\usepackage{slashed}
\usepackage{physics}

\begin{document}

\title{
Warm Inflation with the Standard Model
}
\author{Kim V. Berghaus$^1$}
\author{Marco Drewes$^{2,3}$}
\author{Sebastian Zell$^{4,5,2}$}

\affiliation{$^1$Walter Burke Institute for Theoretical Physics, California Institute of Technology, CA 91125, USA}

\affiliation{$^2$Centre for Cosmolgy, Particle Physics and Phenomenology (CP3), Universit\'e catholique de Louvain, Chemin du Cyclotron 2, B-1348 Louvan-la Neuve, Belgium}

\affiliation{$^3$ Physik–Department, Technische Universität München, James Franck Straße 1, D-85748 Garching, Germany}

\affiliation{$^4$Arnold Sommerfeld Center, Ludwig-Maximilians-Universit\"at, Theresienstraße 37, 80333 M\"unchen, Germany}

\affiliation{$^5$Max-Planck-Institut f\"ur Physik, Boltzmannstr. 8, 85748 Garching b. M\"unchen, Germany}

\begin{abstract}

We show for the first time that warm inflation is feasible with Standard Model (SM) gauge interactions alone. Our model consists of a minimal extension of the SM by a single scalar inflaton field with an axion-like coupling to gluons and a monomial potential. The effects of light fermions, which were previously argued to render warm inflation with the SM impossible, are alleviated by Hubble dilution of their chiral chemical potentials. Our model only features one adjustable combination of parameters and accommodates all inflationary observables. We briefly discuss
implications for axion experiments, dark matter, and the strong CP-problem.

\end{abstract}

\maketitle

\textit{Introduction}---Cosmic inflation \cite{Starobinsky:1980te, Guth:1980zm, Linde:1981mu, Mukhanov:1981xt}, an early period of accelerated cosmic expansion, provides the leading explanation for the observed near-homogeneity and isotropy of our Universe.  However, it remains unknown what mechanism drove inflation, and how it is connected to microphysical theories of particle physics. 
Many models can effectively be parameterised by a single scalar field $\phi$, the potential of which, $V(\phi)$, dominated the cosmic energy budget. However, simple renormalizable potentials with a single parameter have long been ruled out by observations of the cosmic microwave background (CMB) \cite{Planck:2018jri, BICEP:2021xfz}. A plethora of models with more complicated potentials exist~\cite{Martin:2013tda}, but they usually contain more parameters, making an observational discrimination impossible. Moreover, for many models it is not clear how they can be realized microphysically and related to theories of particle physics, in particular the Standard Model (SM). 

Some of these issues can be addressed by \emph{warm inflation} \cite{Berera:1995ie}, a variant of the inflationary paradigm in which dissipative effects maintain the presence of a thermal bath during inflation, see e.g.~\cite{Kamali:2023lzq} for a review. This allows for accelerated
expansion on steeper potentials, alleviating the long-standing difficulty of protecting the flatness of the inflaton potential from quantum corrections~\cite{Copeland:1994vg,Binetruy:1996xj,Dvali:1998pa,Arkani-Hamed:2003wrq,Baumann:2014nda}.
Moreover, thermal fluctuations provide an additional source of scalar perturbations in the metric, effectively suppressing the tensor-to-scalar ratio $r$, and hence rescuing models that would be observationally excluded in their absence. Finally, non-Gaussianities with a unique bispectral shape 
\cite{Bastero-Gil:2014raa, Mirbabayi:2022cbt} provide a pathway towards discovering evidence of inflation.

Arguably, the main challenge of warm inflation has been to embed it in a microphysical theory
\cite{Yokoyama:1998ju}.
The fundamental problem is that, in perturbative computations, the inflaton dissipation coefficient and thermal correction to the effective potential originate from the imaginary and real parts of the same Feynman diagrams,  
making it difficult to enhance the former without introducing sizable corrections to the latter that spoil the flatness of the potential. 
This issue can be avoided if the dissipation arises from an axion-like coupling to the topological winding number in non-Abelian gauge theory, which gives rise to a new form of thermal friction \cite{McLerran:1990de} known as sphaleron heating. 
 It has been shown  that sphaleron heating can indeed support warm inflationary dynamics \cite{Berghaus:2019whh} within a consistent quantum field theory framework \cite{Laine:2021ego},
but previous realizations of this idea generally require new (dark) non-Abelian gauge interactions that serve no other evident purpose. 

Here, we demonstrate that warm inflation can be induced directly by an axion-like coupling between the inflaton and 
quantum chromodynamics (QCD), the strong force of the SM. 
QCD has not previously been considered as a viable candidate for the required non-Abelian force due to the suppression of sphaleron heating by axial chemical potentials, $\mu_A$, that it generates for the quarks  \cite{McLerran:1990de, Berghaus:2020ekh,Berghaus:2024zfg}.  
In the present proposal this problem is alleviated by the Hubble dilution of $\mu_A$ \cite{Drewes:2023khq},
which has been neglected in earlier works but is essential during inflation. 
We construct the first model of warm inflation that is driven by known SM gauge interactions alone. Our proposal is compatible with all observations and only contains two unknown parameters, one in the  quartic inflaton potential and the other one in the coupling to gluons that is potentially experimentally testable.

\textit{Warm Inflation}--- 
We consider inflation to be warm if there is a thermal bath with a temperature $T$ that exceeds the Hubble rate $H$ during a period of accelerated expansion ($T > H$).
To achieve this,
the inflaton field must source particles at a rate that can compensate for their Hubble dilution, and their self-interactions must be fast enough to maintain thermal equilibrium.   
This microphysical process can be captured by an effective description in which a thermal friction coefficient \footnote{See e.g.~\cite{McLerran:1990de,Buldgen:2019dus,Laine:2021ego} and references therein for derivations within quantum field theory.}, 
$\Upsilon$, extracts energy from the slow-rolling inflaton field $\phi$. 
The homogeneous equations of motion of the inflaton expectation value
and radiation density $\rho_R  \equiv \frac{\pi^2}{30} g_* T^4$
read 
\begin{eqnarray}
\label{eq:EOM}
\ddot{\phi} + (3H + \Upsilon) \dot{\phi} + V' &=& 0 \, ,\\
\label{eq:cont}
\dot{\rho}_R + 4 H \rho_R &=& \Upsilon \dot{\phi}^2.
\end{eqnarray}
Here  dots denote derivatives with respect to time, prime denotes derivative with respect to $\phi$, $H \equiv \dot{a}/a$ is the Hubble expansion, with $a$ the scale factor, and $g_*$ corresponds to the number of relativistic degrees of freedom in the thermal bath. 
The cases $Q \equiv \Upsilon/(3H) > 1$ and $Q<1$ are commonly referred to as the strong and weak regime of warm inflation, respectively. 
In both regimes the thermal bath can contribute to sourcing cosmological perturbations, but only in the strong regime thermal friction dominates the slow roll dynamics.

The Friedman equation governs the inflationary dynamics, 
$H^2  = \big(V(\phi) + \rho_R + \frac{1}{2} \dot{\phi}^2 \big)/(3M^2_{\text{pl}}) 
\approx V(\phi)/(3 M_{\text{pl}}^2)$,
where $M_{\text{pl}} = 2.43 \times10^{18} \, \text{GeV}$ is the reduced Planck mass. 
A sustained period of accelerated expansion requires $V(\phi) \gg \rho_R, \dot{\phi}^2$, which can be assured by the usual slow roll conditions, 
$\epsilon_V, \eta_V \ll 1$, 
if the slow roll parameters are modified to be \cite{Berera:2008ar}
\begin{equation}
\label{eq:sr}
\epsilon_V \equiv \frac{M^2_{\text{pl}}}{2(1 +Q)} \left(\frac{V'}{V} \right)^2 , \
\eta_V \equiv \frac{M^2_{\text{pl}}}{(1 +Q)} \frac{V''}{V}.
\end{equation}
From Eq.~\eqref{eq:sr} it is immediately apparent that in the strong regime of warm inflation the slow-roll conditions are more easily satisfied than in its cold counterpart, implying that a given number of $e$-folds,  
\begin{equation}
\label{eq:NCMB}
N_* = \int H dt = \int_{\phi_{\text{end}}}^{\phi_{*}} \frac{1}{M^2_{\text{pl}}} \frac{V}{V'}\left(1 + Q \right) d\phi \, ,
\end{equation}
can be achieved with a smaller field value $\phi_*$ or steeper potential (in the sense of larger $V'/V$).
The slow-roll conditions suppress the first two terms in Eq.~\eqref{eq:EOM} and Eq.~\eqref{eq:cont}, such that 
\begin{equation}
\label{eq:temp}
 \frac{\pi^2}{30} g_{*} T^4  \approx \frac{\Upsilon}{4H} \left(\frac{V'}{(3H + \Upsilon)}\right)^2 
 \, ,
\end{equation}
defines the approximately constant steady-state temperature of the thermal bath $T$.

\textit{Sphaleron heating with the SM}---
Our model is defined by the extension of the SM \footnote{
The inflaton-gluon interaction is mediated by a dimension five operator and therefore must be viewed as an effective field theory (EFT) with a cutoff $\Lambda_{\text{cutoff}} \equiv (\alpha_s/8\pi f)^{-1}$. 
This restricts the validity of our computations to the regime $T < \Lambda_{\text{cutoff}}$.~We verify a posteriori that all temperatures in Table~\ref{tab:N60} lie at least one order of magnitude below this EFT cutoff.
}
\begin{equation}
\label{eq:lag}
\mathcal{L} = \frac{1}{2} \partial^\mu \phi \partial_\mu \phi - V(\phi)  
- \frac{\alpha_s}{8 \pi}\frac{\phi}{f} G^{a}_{\mu \nu} \tilde{G}^{a\mu \nu} \, ,
\end{equation}
where $G^a_{\mu \nu}$ is the QCD field strength tensor, $\tilde{G}^{a\mu\nu} = \frac{1}{2}\epsilon^{\mu\nu\rho\sigma}G^a_{\rho\sigma}$ its dual and $\alpha_s \equiv g^2/4\pi$ with $g$ the QCD gauge coupling. The following considerations, however, hold for a general Yang-Mills SU($N_c$) gauge group.
We first neglect the presence of quarks and include their impact further below.  
 For $T$ larger than the confinement temperature the equation of motion for the expectation value of $\phi$ reads \cite{McLerran:1990de} 
 \begin{equation} \label{inflationEOM}
\ddot{\phi} + 3 H \dot{\phi} + V' = - \frac{\alpha_s}{8 \pi f} \big\langle G^{a}_{\mu \nu} \tilde{G}^{a \mu \nu}  \big \rangle \,.
 \end{equation}
 The thermal average of $G^{a}_{\mu \nu} \tilde{G}^{a \mu \nu}$
 induces a thermal friction coefficient $\Upsilon_{\text{sph}}$ due to real-time sphaleron processes that lead to biased transitions between vacua with different Chern-Simons number $N_{\text{CS}}$, 
 driven by the inflaton velocity $\dot{\phi}$.
The friction coefficient $\Upsilon_{\text{sph}}$
is related to the SU($N_c$) sphaleron rate $ \Gamma_{\text{sph}}\equiv \underset{t \to \infty} {\text{lim}}
\left(N_{\text{CS}}(t) - N_{\text{CS}}(0)\right)^2/(V t) $
(also called the Chern-Simons diffusion rate) via the fluctuation-dissipation theorem \cite{McLerran:1990de,Laine:2016hma,Berghaus:2019whh,Laine:2021ego,Mirbabayi:2022cbt}, 
\begin{equation}
\label{eq:linear}
\frac{\alpha_s}{8 \pi f } \big\langle G^{a}_{\mu \nu} \tilde{G}^{a \mu \nu}  \big \rangle \equiv \Upsilon_{\text{sph}} \dot{\phi}
= \frac{\Gamma_{\text{sph}}}{ 2T f} \frac{\dot{\phi}}{f}.
\end{equation}
By comparing Eqs.~\eqref{inflationEOM} and \eqref{eq:linear} to Eq.~\eqref{eq:EOM}, we can identify $\Upsilon 
= \Upsilon_{\text{sph}}$.
Using the estimate $\Gamma_{\text{sph}} \approx \alpha^5_s N_c^5 T^4$ \cite{Moore:2010jd},  
one finds \cite{Laine:2016hma}
\begin{equation}
\label{eq:sphaleron_heating}
\Upsilon_{\text{sph}} \simeq (\alpha_s N_c)^5  \frac{T^3}{2f^2} \, ,
\end{equation}
where we neglected an additional prefactor of order $1$ with a weak dependence on model parameters and dynamical quantities (see \cite{Moore:2010jd,Laine:2021ego,Laine:2022ytc,Klose:2022rxh}).

The perturbative shift symmetry
of the axion-like coupling in Eq.~\eqref{eq:lag} protects the inflaton potential from thermal corrections \cite{McLerran:1990de,Berghaus:2019whh,Laine:2021ego,Mirbabayi:2022cbt,Klose:2022rxh,Kolesova:2023yfp} that would otherwise endanger the slow-roll dynamics \cite{Yokoyama:1998ju}. 
However, since non-perturbative contributions to $V(\phi)$ from instantons  vanish at temperatures much larger than the QCD confinement scale $\Lambda_{\text{QCD}}$, where $\alpha_s < 1$ (see e.g. \cite{Coleman:1985rnk}), the coupling to QCD cannot generate a potential that drives inflation \footnote{More generally, successful warm inflation in the strong regime requires a potential that breaks the symmetry $\phi \rightarrow \phi + 2 \pi f$ of the axion-like coupling in Eq.~\eqref{eq:lag} \cite{Zell:2024vfn}.}.

The relevant interaction with SM quarks $\psi$ is given by
\begin{equation}
\label{eq:QCD}
\mathcal{L_{\text{QCD}}} 
\supset \sum_f \left(\frac{i}{2}\bar{\psi_f} \slashed D  \psi_f + \text{h.c.}\right) \;,
\end{equation}
where $f$ denotes the sum over SM flavors.
The chiral anomaly of SU($N_c$) leads to a non-conservation of the axial fermion current $j_A^{\mu} = \sum_f \bar{\psi}_f \gamma^\mu \gamma^5 \psi_f $ \cite{McLerran:1990de,Peskin:1995ev},
\begin{equation} \label{anomaly}
\partial_t j_A^{0} = - N_f \Big\langle \frac{\alpha_s G^a_{\mu \nu} \tilde{G}^{a \mu \nu} }{4 \pi} \Big \rangle \, , 
\end{equation}
where the axial charge density $j_A^0$ corresponds to the difference in number density between right-handed and left-handed Dirac fermions. The quarks back-react onto the biased sphaleron transitions such that Eq.~\eqref{eq:linear} changes to \cite{Khlebnikov:1988sr,McLerran:1990de,Giudice:1993bb,Rubakov:1996vz,Moore:1996qs,Moore:1997im,Moore:2010jd}
\begin{equation}
\label{eq:linearplus}
\frac{\alpha_s}{8 \pi f } \big\langle G^{a}_{\mu \nu} \tilde{G}^{a \mu \nu}  \big \rangle 
= \Upsilon_{\text{sph}} \left(\dot{\phi} +\frac{6 f j_A^{0}}{N_c T^2}  \right)  \, .
\end{equation}
In particular, Eqs.~\eqref{anomaly} and \eqref{eq:linearplus} imply that in the absence of a rolling axion, sphalerons relax the axial charge as  
$\partial_t j_A^{0} = - 6 (N_f/N_c) (\Gamma_{\text{sph}}/T^3) j_A^{0}$ \cite{Khlebnikov:1988sr,Giudice:1993bb,Rubakov:1996vz,Moore:1996qs,Moore:1997im,Moore:2010jd}.

We define a chemical potential $\mu_A$ via
\begin{align}
j_A^0 &=  2 N_c N_f  \int \frac{d^3 p}{(2 \pi)^3} \left(\frac{1}{e^{(p-\mu_A)/T} +1} -\frac{1}{e^{(p+\mu_A)/T} +1} \right) \nonumber \\
& = \frac{N_c N_f \mu_A T^2}{3} \;, \label{eq:jOfMu}
\end{align}
where $N_f$ denote the number of flavors, and we have taken the $\mu_A \ll T$ limit in the second line \footnote
{In terms of this chemical potential, 
we see that $\partial_t j_A^{0}$ is independent of $N_c$ but scales with $N_f^2 \mu_A$. 
A heuristic explanation of this scaling is that the production rate of fermions, which is caused by sphaleron transitions and violates chiral symmetry, is proportional to the free energy $N_f \mu_A$ liberated by a sphaleron event and, in addition, each event changes $j_A^{0}$ by $N_f$ \cite{Moore:1997im}.
}.
Using
Eq.~\eqref{eq:linearplus},
Eq.~\eqref{inflationEOM} and the anomaly relation \eqref{anomaly}, we obtain the coupled evolution of the inflaton and the axial chemical potential for a quasi-steady state temperature in an expanding universe \cite{McLerran:1990de} 
\begin{equation} \label{eq:inflatonbackr}
\ddot{\phi} +3H \dot{\phi} + V' = - \Upsilon_{\text{sph}} \left( \dot{\phi} + 2 N_f f \mu_A \right) \, ,
\end{equation}
\begin{equation} \label{eq:eomMu}
\dot{\mu}_A + 3H \mu_A = -\frac{6 f  \Upsilon_{\text{sph}}}{N_c T^2}  \left(\dot{\phi} + 2 N_f f \mu_A  \right) \, . 
\end{equation}
Importantly, here we also take into account Hubble dilution of the chemical potential in the second line. 
When Hubble friction is neglected, there is no solution where $\ddot{\phi}$ vanishes asymptotically \cite{McLerran:1990de}, hence  slow-roll cannot be sustained   \cite{Berghaus:2019whh,Berghaus:2020ekh}. 
In a quasi-static situation, the effect of $H$ can be incorporated in Eq.~\eqref{eq:inflatonbackr} by 
absorbing the term $\Upsilon_{\text{sph}} 2 N_f f \mu_A $ into an effective dissipation coefficient \cite{Drewes:2023khq},
\begin{equation} \label{eq:YEff}
	\Upsilon_{\text{eff}} = \Upsilon_{\text{sph}} \Big/\left(1 + \frac{\Upsilon_{\text{sph}}}{3H} \frac{N_f}{N_c} \frac{12 f^2}{T^2} \right) .
\end{equation}
Sphaleron heating is fully restored if 
$ H/\Upsilon_{\text{eff}} \gg 4 f^2 N_f/(T^2 N_c)$
i.e., the chemical potential is diluted faster than it builds up due to sphaleron transitions. 
Achieving this is generically easier in the weak regime of warm inflation where $\Upsilon_{\text{sph}}<3H$.
Using the Friedmann equation and Eq.~\eqref{eq:sphaleron_heating}, we get
\begin{align}
	\label{eq:Yeffsimple}
	\Upsilon_{\text{eff}} & = 
	N_c^5 \alpha^5_s \frac{T^3}{2f^2} \Big/\left(1+\frac{2 N_f N_c^4 \alpha^5_s T}{ \sqrt{V(\phi)/(3M^2_{\text{pl}}})}\right) \, .
\end{align}
Note that Eq.~\eqref{eq:Yeffsimple} can yield different functional dependencies of
$\Upsilon_{\text{eff}}$ on $\phi$ and $T$ if the second term dominates the denominator, depending on the shape of $V(\phi)$, 
\begin{equation}\label{DampingDependence}
	\Upsilon_{\text{eff}} \propto \frac{T^2}{f^2} \frac{\sqrt{V(\phi)}} {M_{\text{pl}}} \, .
\end{equation}
In addition to the gauge interactions in \eqref{eq:QCD} quarks also have Yukawa couplings, $y_f H \bar{\psi_f} \psi_f $,
which mediate chirality violation processes 
either by generating fermion masses \cite{Boyarsky:2020cyk} or through thermal scatterings \cite{Bodeker:2019ajh} \footnote{Moreover, SU$(2)$ sphalerons provide an additional source of chirality violation \cite{Drewes:2023khq}, but this effect turns out to be negligible since $(N_c \alpha_s)^5 \gg (2\alpha_w)^5$, where $\alpha_w$ is the weak fine structure constant.}.
Both make the SM flavors distinguishable.
An equation of motion for each chemical potential $\mu_{A_f}$ can be obtained from \eqref{eq:eomMu} by omitting the factor $N_f$ and adding a term $- \Gamma_{\text{ch} f} \mu_A$ on the r.h.s., which can effectively be incorporated into $\Upsilon_{\text{eff}}$ by expressing the factor $N_f$ terms of a sum over flavors with the replacement $3H \to 3H + \Gamma_{\text{ch} f}$. 
If all $\Gamma_{\text{ch} f}$ are either much smaller or much larger than $H$ and $\Upsilon_{\text{sph}}$, one can simply continue using the unflavored equation \eqref{eq:YEff}, but set $N_f$ to the number of SM flavors for which $\Gamma_{\text{ch} f} \ll \Upsilon_{\text{sph}}$, which for $\Gamma_{\text{ch} f} \sim 10^{-2} y_f^2 T$ \cite{Bodeker:2019ajh} amounts to $ y_f^2 \lesssim 10^3 \alpha^5_s N_c^4$. This condition is only violated for the top quark, hence we use \eqref{eq:YEff} with $N_f=5$ \footnote
{For the bottom quark $\Gamma_{\text{ch} f} \simeq \Upsilon_{\text{sph}}$, provided $\alpha_s$ runs to sufficiently low values.}.
We further set $N_c = 3$ and obtain $\alpha_s$ from running to the energy scale $T$ with the known SM $\beta$-functions \cite{Sartore:2020gou} \footnote{Thermal corrections to the vacuum $\beta$-functions are expected to be negligible at temperatures greatly exceeding the QCD confinement scale  \cite{Kaczmarek:2005ui,Braun:2005uj}.}.

\begin{figure}
	\includegraphics[width=0.5\textwidth]{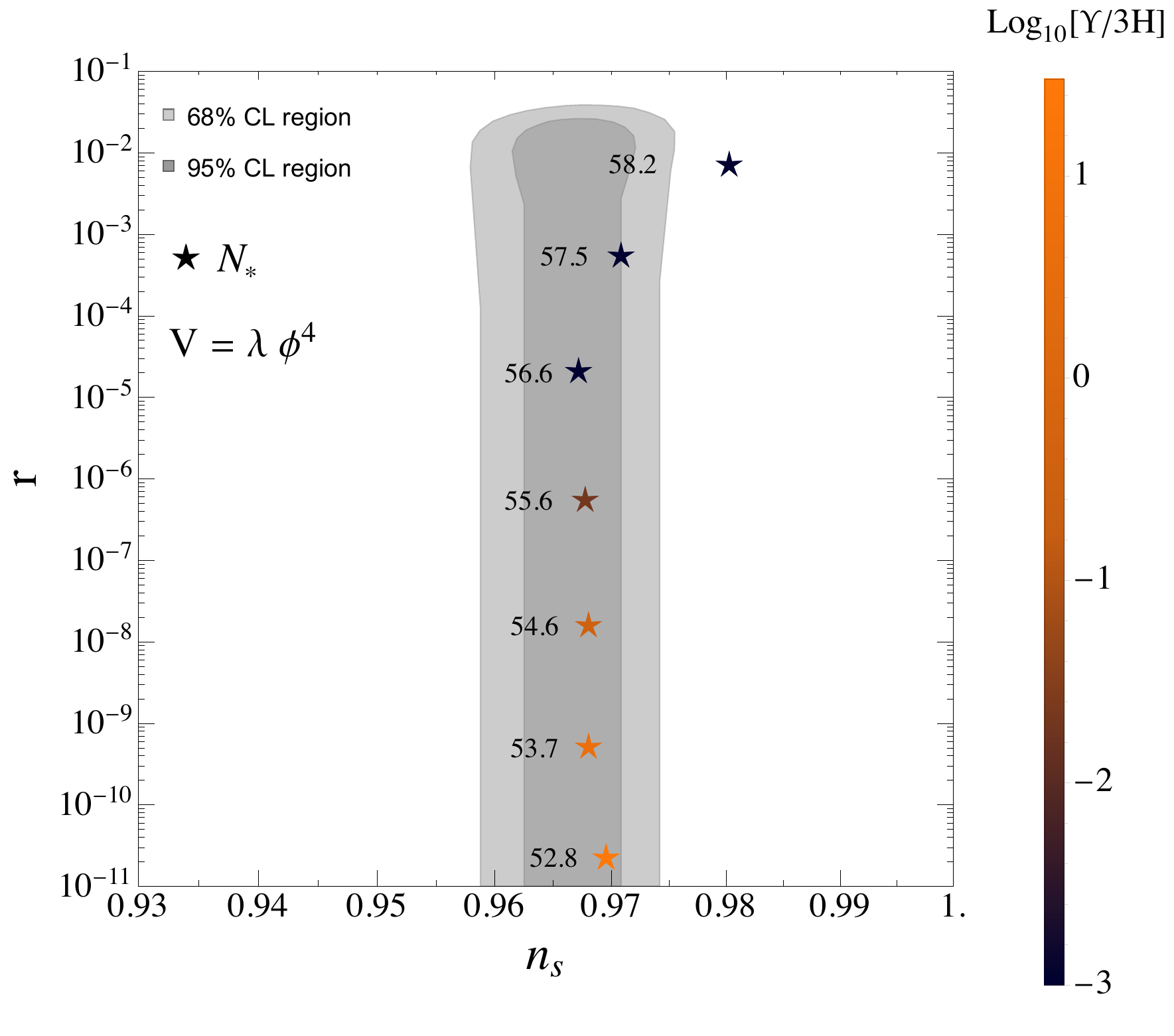} 
	\caption{The predictions of warm inflation with the SM
    for a quartic inflaton potential.
    The allowed 68$\%$ and 95$\%$ contour regions correspond to the constraints 
from the combined analysis of Planck \cite{Planck:2018jri} and BICEP/Keck \cite{BICEP:2021xfz} data. 
    Proposed and upcoming CMB experiments
    will improve this bound by about a factor of four (AliCPT \cite{Li:2017drr,Liu:2025sut}, Simons Observatory \cite{SimonsObservatory:2018koc}, Simons Array \cite{POLARBEAR:2015ixw}) to twenty (LiteBIRD \cite{LiteBIRD:2022cnt}, CMB-S4 \cite{CMB-S4:2020lpa}).} 
    \label{fig:rvsns}
\end{figure}

\textit{Working Model}---
We choose the quartic potential 
\begin{equation}\label{lambda4}
V(\phi) = \lambda \phi^4 \, ,
\end{equation}
to fully determine the Lagrangian \eqref{eq:lag} with two unknown model parameters $f$ and  $\lambda$. 
Matching the observed  amplitude $A_s(k_*) \approx 2 \times 10^{-9}$ of scalar perturbations $\Delta^2_s(k) = A_s(k_*) \left(k/k_* \right)^{n_s(k_*)-1}$ at the pivot scale $k_* = 0.05 \, \text{Mpc}^{-1}$ \cite{Planck:2018jri} establishes a relation between them, effectively leaving us with one single free parameter.
Practically, we use Eqs.~\eqref{eq:NCMB} and \eqref{eq:temp}, with $\Upsilon = \Upsilon_{\text{eff}}$ and \eqref{lambda4} to numerically solve for the temperature, $T$, and subsequently $\Upsilon_{\text{eff}}$ as a function of $\phi$, $\lambda$, and $f$.  
Varying $\lambda$ scans along values of $Q=\Upsilon_{\text{eff}}/(3H)$. 
With this and the approximation of instantaneous reheating we can compute $N_*$, $A_s$, $n_s$, and the tensor-to-scalar ratio, $r \equiv \Delta^2_h/\Delta^2_s$ from the relations given in the Appendix.
Since our model effectively only has one free parameter, we predict a line in the $n_s$-$r$-plane, as displayed in Figure \ref{fig:rvsns}. 

We find that our model is compatible with observations in both the strong regime  and the weak regime, 
consistent with what was previously found for dark gauge groups \cite{Mirbabayi:2022cbt}.
For $Q\gg 1$ 
the denominator in Eq.~\eqref{eq:Yeffsimple} suppresses the efficiency of sphaleron heating ($\Upsilon_{\rm eff} \ll \Upsilon_{\rm sph}$), 
and $\Upsilon_{\text{eff}}$ 
follows the scaling in Eq.~\eqref{DampingDependence}.
We do not compare to CMB observables in this regime as we can no longer compute perturbations reliably with the formulae given in the appendix. 
This suppression is, however, already alleviated for $Q \lesssim \mathcal{O}(10)$, i.e., within the strong regime.
Column 7 in Table~\ref{tab:N60} shows that the deviation of \eqref{eq:Yeffsimple} from $\Upsilon_{\text{sph}}$ amounts to less than $10\%$. 
For smaller $Q$ the restoration becomes increasingly efficient, so that $\Upsilon_{\rm eff}$ further approaches $ \Upsilon_{\rm sph}$ in the weak regime. 
Thermally produced cosmological perturbations become negligible 
for $Q < 10^{-4}$, practically recovering standard cold inflation.
The predicted warm inflation temperatures lie below $
 T \lesssim 6 \times 10^{14}  \, \text{GeV},$ with temperatures decreasing for larger values of $Q$. The corresponding quartic couplings for which inflation is warm lie below $\lambda \lesssim 10^{-13}$.
We provide a comprehensive list of all the relevant parameters discussed here 
in Table \ref{tab:N60}. 

In our model, the interaction of the inflaton $\phi$ with the SM is already fully specified. Therefore, the transition from inflation to radiation-domination is calculable,  
and in the strong regime, accelerated expansion directly evolves to a period dominated by a quark-gluon plasma. 
This is reflected in the fact that we can uniquely predict $r$ as a function of $n_s$ (see Figure \ref{fig:rvsns}), with no additional uncertainty arising from unknown interactions of the inflaton with the SM.
Moreover, the direct coupling of the inflaton to QCD can lead to observable signatures 
at low energies \footnote{In contrast, the connection between cold inflation and particle physics can only be probed indirectly \cite{Drewes:2019rxn,Drewes:2022nhu}, with the notable exception of Higgs inflation \cite{Bezrukov:2007ep} (see e.g.~\cite{Shaposhnikov:2020fdv}).}. The values for the coupling $f\sim 10^{10} -10^{13} \, \text{GeV}$ lie within reach of proposed searches for axion-like particles \cite{Antel:2023hkf,AxionLimits}. 
However, Hubble dilution alone suffices to ensure that $\phi \ll f$ at the temperature of the QCD phase transition, implying that $\phi$-oscillations around the minimum of the potential $V(\phi) \simeq \lambda \phi^4 + \Lambda_{\text{QCD}}^4 \left(1-\cos\left(\phi/f-\theta\right)\right)$ only contribute a small fraction to the Dark Matter density,
where 
$\theta$ is the CP-violating angle in QCD. 
Since $\lambda \phi^4$ still gives an important contribution to the potential at low energies, satisfying the experimental bound on CP-violation in the strong interaction requires $|\phi/f-\theta|\ll 10^{-10}$ at the minimum, i.e., the same amount of tuning as in QCD, and can potentially be achieved with a standard QCD-axion \cite{Peccei:1977hh,Weinberg:1977ma,Wilczek:1977pj}.

\textit{Conclusion}---
We showed for the first time that warm inflation can be driven by sphaleron heating from SM gauge interactions alone. Our model comprises a minimal extension of the SM by a single scalar inflaton field $\phi$ with an axion-like coupling to gluons \eqref{eq:lag}.
A crucial point is that the suppressing effect which light fermions have on the efficiency of sphaleron heating is partially alleviated by the expansion of the universe, leading to the effective friction coefficient in Eq.~\eqref{eq:YEff}.
Notably, this coefficient can have different functional dependencies on the inflaton field value 
for different effective potentials in Eq.~\eqref{DampingDependence}.
Additionally, it predicts a smooth transition from a warm inflationary period to  
a Universe dominated by a quark-gluon plasma. 
For a monomial quartic potential \eqref{lambda4} the model only contains two parameters and is consistent with CMB observations as well as laboratory experiments. Its minimality in terms of both, the potential during inflation and the clear connection to the SM, make this scenario highly predictive and a target for tests with future CMB and large scale structure surveys as well as laboratory experiments, offering a unique opportunity to probe the connection between cosmic inflation and particle physics.

\textit{Note added}---
Recently, CMB predictions of our model were further investigated in \cite{ORamos:2025uqs,Laine:2025rll}. Both works confirm that our proposal is compatible with current observational bounds.

\textit{Acknowledgments}---We are grateful to Mikko Laine, Alica Rogelj, and Mikhail Shaposhnikov for their valuable comments on the manuscript.
K.B. thanks the U.S. Department of Energy, Office of Science, Office of High Energy Physics, under Award Number DE-SC0011632 and the Walter Burke Institute for Theoretical Physics.
The work of M.D. was partially funded by the Deutsche Forschungsgemeinschaft (DFG, German Research Foundation) - SFB 1258 - 283604770.
The work of S.Z.~was supported by the European Research Council Gravites Horizon Grant AO number: 850 173-6.
\\
\textit{Disclaimer:} Funded by the European Union.
Views and opinions expressed are however those of
the authors only and do not necessarily reflect those of the European Union or European Research Council. 

\begin{appendix}

\begin{table*}[t]
\centering
\begin{tabular}{|c||c|c|||c|c|c|c|c|c|c|c|c|}
\hline
$\lambda$ & $\frac{\phi_*}{M_{\text{pl}}}$ & f [$10^{12}\,$GeV] & T [$10^{12}\,$GeV] &$\alpha_s$  & T/H & $\frac{2 N_f N^4_c \alpha^5_s T}{\sqrt{\lambda \phi^4_* /3 M^2_{\text{pl}}}}$ & $T/\Lambda_{\text{cutoff}}$ & $N_*$ & Q & $n_s$& r  \\
\hline
\hline
$10^{-21}$ &5.02 & $0.146$ & $8.75$ & 0.0269  &7820& $8.99 \times 10^{-2}$ & $0.064$ &52.8 &$14.9$ & 0.9696 & $2.19 \times 10^{-11}$ \\
\hline
$10^{-20}$ & 6.19 & $0.264$ & $19.1$ & 0.0263  & 3551 & $3.63 \times 10^{-2}$ & $0.076 $ &  53.7 &$9.22$ &0.9681 & $5.02 \times 10^{-10}$ \\
\hline
$10^{-19}$ & 8.22 & $0.518$ & $44.1$ & 0.0257 &1474 & $1.34 \times 10^{-2}$ &  $0.087$& 54.6&$4.78$&0.9681 &$1.54 \times 10^{-8}$ \\
\hline
$10^{-18}$ & 11.1 & $1.05$ & $101$ & 0.0251 &582 & $4.69 \times 10^{-3}$ & $0.097$& 55.6 &$2.19$ & 0.9678 &$5.29 \times 10^{-7} $\\
\hline
$10^{-17}$ &15.6  & $2.30$ & $219$ & 0.0246& 202 & $1.46 \times 10^{-3}$ & 0.093 &56.6 &$0.661$ & 0.9672 & $2.01 \times 10^{-5}$ \\
\hline
$10^{-16}$ & 20.0  & $4.95$ & $329$ & 0.0243 & 58.9 & $4.03 \times 10^{-4}$& $0.064$& 57.5 &$ 0.0889 $&0.9709 &$5.34 \times 10^{-4}$ \\
\hline
$10^{-15}$ &21.4  & $8.71$ & $381$ & 0.0242 & 18.8 & $1.26 \times 10^{-4}$ & $0.042 $&58.2 &$0.0122 $& 0.9803 & $6.96\times 10^{-3}$ \\
\hline
\end{tabular}
\caption{Parameters of interest for warm SM inflation. $\lambda$, the potential slope, is a free parameter and scans the parameter space along $Q$. $\phi_*$  and $f$ are fixed by
$N_*$,
and $A_s \approx 2 \times 10^{-9} $.
All other parameters are derived. 
The EFT cutoff is $\Lambda_{\text{cutoff}} \equiv (\alpha_s/8\pi f)^{-1}$.
We present values to three significant figures to facilitate reproducibility when using  Eq.~\eqref{eq:sphaleron_heating} and Eq.~\eqref{eq:As}. 
However, when accounting for theoretical uncertainties, we estimate that only one significant figure can be considered as reliable for the derived quantities (two figures for $n_s$, theoretical uncertainty not shown in Fig.~\ref{fig:rvsns}).}.
\label{tab:N60}
\end{table*}

\textit{Appendix---}
The spectrum of cosmic perturbations for warm inflation has not been derived within quantum field theory at the same level of rigor as for standard inflation, and the predictions presented in the literature exhibit minor discrepancies \cite{Graham:2009bf,Bastero-Gil:2011rva,Mirbabayi:2022cbt, Montefalcone:2023pvh,Berghaus:2024zfg,Laine:2024wyv,Rodrigues:2025neh,ORamos:2025uqs,Laine:2025rll}.
However, our main conclusions are not sensitive to these differences. We use the predictions for $A_s$ from \cite{Mirbabayi:2022cbt} (recently confirmed by \cite{Laine:2025rll}), 
\begin{equation}
\label{eq:scalar}
\Delta^2_s(\phi_*) = \frac{H^2(\phi_*)}{4\pi^2 \dot{\phi}^2(\phi_*)} \left({H(\phi_*)^2} + H(\phi_*) T(\phi_*) F(Q(\phi_*)) \right) \, ,
\end{equation}
where \cite{Mirbabayi:2022cbt,Berghaus:2024zfg}
\begin{align}
\label{eq:As}
F(Q_*) & = 3.24 \times 10^{-4} Q_*^7 \nonumber \\
& + 168 Q_*\left(\frac{1}{3}\left(1 +\frac{9Q_*^2}{25}  \right) +\frac{2}{3} \tanh \frac{1}{30Q_*} \right) \,.
\end{align} 
The resulting scalar index, $n_s-1 \equiv d\ln \Delta^2_s/dN$, cannot be written in closed form in all generality for the regime in which Q transitions from the weak to the strong regime, and we thus calculate it numerically.

The scalar-to-tensor ratio is
\begin{align}
r = \frac{8 \dot{\phi^2_*}}{M^2_{\text{pl}} H_* T_* F(Q_*)} \, .
\end{align} 
Eq.~\eqref{eq:As} was derived for $ \Upsilon_{\text{sph}}$, rather than  $\Upsilon_{\text{eff}}$ for which the temperature no longer scales as a simple power law. 
This is a subtle complication as it has been shown that the scalar perturbations are quite sensitive to the scaling with temperature in the friction coefficient for $Q > 1$ \cite{Graham:2009bf,Bastero-Gil:2011rva}. 
However, in the absence of calculations of the scalar power spectrum given $\Upsilon_{\text{eff}}$, we  proceed with Eq.~\eqref{eq:scalar} as the best available approximation, and column 7 in table \ref{tab:N60} confirms a posteriori that the error is small.

Finally, the number of inflationary e-foldings after the generation of CMB perturbations is (c.f.~\cite{Rubio:2019ypq})
\begin{equation} \label{NReheating}
    N_* = \ln \left(\left(\frac{g_{*0}}{g_*}\right)^{1/3} \frac{T_0}{T_{RH}}\frac{H_*}{k_*}\right) \;,
\end{equation}
to be matched with Eq.~\eqref{eq:NCMB}. Here $g_{*0}=3.94$ \cite{Husdal:2016haj} is the effective number of degrees of freedom today, 
$T_0\approx 2.7\,\text{K}$ corresponds to the present CMB temperature, $H_*$ is the Hubble scale at CMB generation, and we take $k_*=0.05 \, \text{Mpc}^{-1}$ as pivot scale. 
Given that the inflaton interactions are fully specified in our model, reheating is in principle calculable. In Eq.~\eqref{NReheating}, we have taken the transition from inflation to radiation domination as instantaneous and believe this to be a good approximation due to the strong coupling $\sim \alpha T/f$
with SM degrees of freedom. Consequently, we set $T_{RH}$ to the temperature of the thermal plasma 
when $\epsilon_V(\phi_{\text{end}}) =1$.

\end{appendix}
\bibliography{WI}{}

\begin{thebibliography}{67}%
\makeatletter
\providecommand \@ifxundefined [1]{%
 \@ifx{#1\undefined}
}%
\providecommand \@ifnum [1]{%
 \ifnum #1\expandafter \@firstoftwo
 \else \expandafter \@secondoftwo
 \fi
}%
\providecommand \@ifx [1]{%
 \ifx #1\expandafter \@firstoftwo
 \else \expandafter \@secondoftwo
 \fi
}%
\providecommand \natexlab [1]{#1}%
\providecommand \enquote  [1]{``#1''}%
\providecommand \bibnamefont  [1]{#1}%
\providecommand \bibfnamefont [1]{#1}%
\providecommand \citenamefont [1]{#1}%
\providecommand \href@noop [0]{\@secondoftwo}%
\providecommand \href [0]{\begingroup \@sanitize@url \@href}%
\providecommand \@href[1]{\@@startlink{#1}\@@href}%
\providecommand \@@href[1]{\endgroup#1\@@endlink}%
\providecommand \@sanitize@url [0]{\catcode `\\12\catcode `\$12\catcode `\&12\catcode `\#12\catcode `\^12\catcode `\_12\catcode `\%12\relax}%
\providecommand \@@startlink[1]{}%
\providecommand \@@endlink[0]{}%
\providecommand \url  [0]{\begingroup\@sanitize@url \@url }%
\providecommand \@url [1]{\endgroup\@href {#1}{\urlprefix }}%
\providecommand \urlprefix  [0]{URL }%
\providecommand \Eprint [0]{\href }%
\providecommand \doibase [0]{http://dx.doi.org/}%
\providecommand \selectlanguage [0]{\@gobble}%
\providecommand \bibinfo  [0]{\@secondoftwo}%
\providecommand \bibfield  [0]{\@secondoftwo}%
\providecommand \translation [1]{[#1]}%
\providecommand \BibitemOpen [0]{}%
\providecommand \bibitemStop [0]{}%
\providecommand \bibitemNoStop [0]{.\EOS\space}%
\providecommand \EOS [0]{\spacefactor3000\relax}%
\providecommand \BibitemShut  [1]{\csname bibitem#1\endcsname}%
\let\auto@bib@innerbib\@empty
\bibitem [{\citenamefont {Starobinsky}(1980)}]{Starobinsky:1980te}%
  \BibitemOpen
  \bibfield  {author} {\bibinfo {author} {\bibfnamefont {Alexei~A.}\ \bibnamefont {Starobinsky}},\ }\bibfield  {title} {\enquote {\bibinfo {title} {{A New Type of Isotropic Cosmological Models Without Singularity}},}\ }\href {\doibase 10.1016/0370-2693(80)90670-X} {\bibfield  {journal} {\bibinfo  {journal} {Phys. Lett. B}\ }\textbf {\bibinfo {volume} {91}},\ \bibinfo {pages} {99--102} (\bibinfo {year} {1980})}\BibitemShut {NoStop}%
\bibitem [{\citenamefont {Guth}(1981)}]{Guth:1980zm}%
  \BibitemOpen
  \bibfield  {author} {\bibinfo {author} {\bibfnamefont {Alan~H.}\ \bibnamefont {Guth}},\ }\bibfield  {title} {\enquote {\bibinfo {title} {{The Inflationary Universe: A Possible Solution to the Horizon and Flatness Problems}},}\ }\href {\doibase 10.1103/PhysRevD.23.347} {\bibfield  {journal} {\bibinfo  {journal} {Phys. Rev. D}\ }\textbf {\bibinfo {volume} {23}},\ \bibinfo {pages} {347--356} (\bibinfo {year} {1981})}\BibitemShut {NoStop}%
\bibitem [{\citenamefont {Linde}(1982)}]{Linde:1981mu}%
  \BibitemOpen
  \bibfield  {author} {\bibinfo {author} {\bibfnamefont {Andrei~D.}\ \bibnamefont {Linde}},\ }\bibfield  {title} {\enquote {\bibinfo {title} {{A New Inflationary Universe Scenario: A Possible Solution of the Horizon, Flatness, Homogeneity, Isotropy and Primordial Monopole Problems}},}\ }\href {\doibase 10.1016/0370-2693(82)91219-9} {\bibfield  {journal} {\bibinfo  {journal} {Phys. Lett. B}\ }\textbf {\bibinfo {volume} {108}},\ \bibinfo {pages} {389--393} (\bibinfo {year} {1982})}\BibitemShut {NoStop}%
\bibitem [{\citenamefont {Mukhanov}\ and\ \citenamefont {Chibisov}(1981)}]{Mukhanov:1981xt}%
  \BibitemOpen
  \bibfield  {author} {\bibinfo {author} {\bibfnamefont {Viatcheslav~F.}\ \bibnamefont {Mukhanov}}\ and\ \bibinfo {author} {\bibfnamefont {G.~V.}\ \bibnamefont {Chibisov}},\ }\bibfield  {title} {\enquote {\bibinfo {title} {{Quantum Fluctuations and a Nonsingular Universe}},}\ }\href@noop {} {\bibfield  {journal} {\bibinfo  {journal} {JETP Lett.}\ }\textbf {\bibinfo {volume} {33}},\ \bibinfo {pages} {532--535} (\bibinfo {year} {1981})}\BibitemShut {NoStop}%
\bibitem [{\citenamefont {Akrami}\ \emph {et~al.}(2020)\citenamefont {Akrami} \emph {et~al.}}]{Planck:2018jri}%
  \BibitemOpen
  \bibfield  {author} {\bibinfo {author} {\bibfnamefont {Y.}~\bibnamefont {Akrami}} \emph {et~al.} (\bibinfo {collaboration} {Planck}),\ }\bibfield  {title} {\enquote {\bibinfo {title} {{Planck 2018 results. X. Constraints on inflation}},}\ }\href {\doibase 10.1051/0004-6361/201833887} {\bibfield  {journal} {\bibinfo  {journal} {Astron. Astrophys.}\ }\textbf {\bibinfo {volume} {641}},\ \bibinfo {pages} {A10} (\bibinfo {year} {2020})},\ \Eprint {http://arxiv.org/abs/1807.06211} {arXiv:1807.06211 [astro-ph.CO]} \BibitemShut {NoStop}%
\bibitem [{\citenamefont {Ade}\ \emph {et~al.}(2021)\citenamefont {Ade} \emph {et~al.}}]{BICEP:2021xfz}%
  \BibitemOpen
  \bibfield  {author} {\bibinfo {author} {\bibfnamefont {P.~A.~R.}\ \bibnamefont {Ade}} \emph {et~al.} (\bibinfo {collaboration} {BICEP, Keck}),\ }\bibfield  {title} {\enquote {\bibinfo {title} {{Improved Constraints on Primordial Gravitational Waves using Planck, WMAP, and BICEP/Keck Observations through the 2018 Observing Season}},}\ }\href {\doibase 10.1103/PhysRevLett.127.151301} {\bibfield  {journal} {\bibinfo  {journal} {Phys. Rev. Lett.}\ }\textbf {\bibinfo {volume} {127}},\ \bibinfo {pages} {151301} (\bibinfo {year} {2021})},\ \Eprint {http://arxiv.org/abs/2110.00483} {arXiv:2110.00483 [astro-ph.CO]} \BibitemShut {NoStop}%
\bibitem [{\citenamefont {Martin}\ \emph {et~al.}(2014)\citenamefont {Martin}, \citenamefont {Ringeval},\ and\ \citenamefont {Vennin}}]{Martin:2013tda}%
  \BibitemOpen
  \bibfield  {author} {\bibinfo {author} {\bibfnamefont {Jerome}\ \bibnamefont {Martin}}, \bibinfo {author} {\bibfnamefont {Christophe}\ \bibnamefont {Ringeval}}, \ and\ \bibinfo {author} {\bibfnamefont {Vincent}\ \bibnamefont {Vennin}},\ }\bibfield  {title} {\enquote {\bibinfo {title} {{Encyclop\ae{}dia Inflationaris}: {Opiparous Edition}},}\ }\href {\doibase 10.1016/j.dark.2024.101653} {\bibfield  {journal} {\bibinfo  {journal} {Phys. Dark Univ.}\ }\textbf {\bibinfo {volume} {5-6}},\ \bibinfo {pages} {75--235} (\bibinfo {year} {2014})},\ \Eprint {http://arxiv.org/abs/1303.3787} {arXiv:1303.3787 [astro-ph.CO]} \BibitemShut {NoStop}%
\bibitem [{\citenamefont {Berera}(1995)}]{Berera:1995ie}%
  \BibitemOpen
  \bibfield  {author} {\bibinfo {author} {\bibfnamefont {Arjun}\ \bibnamefont {Berera}},\ }\bibfield  {title} {\enquote {\bibinfo {title} {{Warm inflation}},}\ }\href {\doibase 10.1103/PhysRevLett.75.3218} {\bibfield  {journal} {\bibinfo  {journal} {Phys. Rev. Lett.}\ }\textbf {\bibinfo {volume} {75}},\ \bibinfo {pages} {3218--3221} (\bibinfo {year} {1995})},\ \Eprint {http://arxiv.org/abs/astro-ph/9509049} {arXiv:astro-ph/9509049} \BibitemShut {NoStop}%
\bibitem [{\citenamefont {Kamali}\ \emph {et~al.}(2023)\citenamefont {Kamali}, \citenamefont {Motaharfar},\ and\ \citenamefont {O.~Ramos}}]{Kamali:2023lzq}%
  \BibitemOpen
  \bibfield  {author} {\bibinfo {author} {\bibfnamefont {Vahid}\ \bibnamefont {Kamali}}, \bibinfo {author} {\bibfnamefont {Meysam}\ \bibnamefont {Motaharfar}}, \ and\ \bibinfo {author} {\bibfnamefont {Rudnei}\ \bibnamefont {O.~Ramos}},\ }\bibfield  {title} {\enquote {\bibinfo {title} {{Recent Developments in Warm Inflation}},}\ }\href {\doibase 10.3390/universe9030124} {\bibfield  {journal} {\bibinfo  {journal} {Universe}\ }\textbf {\bibinfo {volume} {9}},\ \bibinfo {pages} {124} (\bibinfo {year} {2023})},\ \Eprint {http://arxiv.org/abs/2302.02827} {arXiv:2302.02827 [hep-ph]} \BibitemShut {NoStop}%
\bibitem [{\citenamefont {Copeland}\ \emph {et~al.}(1994)\citenamefont {Copeland}, \citenamefont {Liddle}, \citenamefont {Lyth}, \citenamefont {Stewart},\ and\ \citenamefont {Wands}}]{Copeland:1994vg}%
  \BibitemOpen
  \bibfield  {author} {\bibinfo {author} {\bibfnamefont {Edmund~J.}\ \bibnamefont {Copeland}}, \bibinfo {author} {\bibfnamefont {Andrew~R.}\ \bibnamefont {Liddle}}, \bibinfo {author} {\bibfnamefont {David~H.}\ \bibnamefont {Lyth}}, \bibinfo {author} {\bibfnamefont {Ewan~D.}\ \bibnamefont {Stewart}}, \ and\ \bibinfo {author} {\bibfnamefont {David}\ \bibnamefont {Wands}},\ }\bibfield  {title} {\enquote {\bibinfo {title} {{False vacuum inflation with Einstein gravity}},}\ }\href {\doibase 10.1103/PhysRevD.49.6410} {\bibfield  {journal} {\bibinfo  {journal} {Phys. Rev. D}\ }\textbf {\bibinfo {volume} {49}},\ \bibinfo {pages} {6410--6433} (\bibinfo {year} {1994})},\ \Eprint {http://arxiv.org/abs/astro-ph/9401011} {arXiv:astro-ph/9401011} \BibitemShut {NoStop}%
\bibitem [{\citenamefont {Binetruy}\ and\ \citenamefont {Dvali}(1996)}]{Binetruy:1996xj}%
  \BibitemOpen
  \bibfield  {author} {\bibinfo {author} {\bibfnamefont {P.}~\bibnamefont {Binetruy}}\ and\ \bibinfo {author} {\bibfnamefont {G.~R.}\ \bibnamefont {Dvali}},\ }\bibfield  {title} {\enquote {\bibinfo {title} {{D term inflation}},}\ }\href {\doibase 10.1016/S0370-2693(96)01083-0} {\bibfield  {journal} {\bibinfo  {journal} {Phys. Lett. B}\ }\textbf {\bibinfo {volume} {388}},\ \bibinfo {pages} {241--246} (\bibinfo {year} {1996})},\ \Eprint {http://arxiv.org/abs/hep-ph/9606342} {arXiv:hep-ph/9606342} \BibitemShut {NoStop}%
\bibitem [{\citenamefont {Dvali}\ and\ \citenamefont {Tye}(1999)}]{Dvali:1998pa}%
  \BibitemOpen
  \bibfield  {author} {\bibinfo {author} {\bibfnamefont {G.~R.}\ \bibnamefont {Dvali}}\ and\ \bibinfo {author} {\bibfnamefont {S.~H.~Henry}\ \bibnamefont {Tye}},\ }\bibfield  {title} {\enquote {\bibinfo {title} {{Brane inflation}},}\ }\href {\doibase 10.1016/S0370-2693(99)00132-X} {\bibfield  {journal} {\bibinfo  {journal} {Phys. Lett. B}\ }\textbf {\bibinfo {volume} {450}},\ \bibinfo {pages} {72--82} (\bibinfo {year} {1999})},\ \Eprint {http://arxiv.org/abs/hep-ph/9812483} {arXiv:hep-ph/9812483} \BibitemShut {NoStop}%
\bibitem [{\citenamefont {Arkani-Hamed}\ \emph {et~al.}(2003)\citenamefont {Arkani-Hamed}, \citenamefont {Cheng}, \citenamefont {Creminelli},\ and\ \citenamefont {Randall}}]{Arkani-Hamed:2003wrq}%
  \BibitemOpen
  \bibfield  {author} {\bibinfo {author} {\bibfnamefont {Nima}\ \bibnamefont {Arkani-Hamed}}, \bibinfo {author} {\bibfnamefont {Hsin-Chia}\ \bibnamefont {Cheng}}, \bibinfo {author} {\bibfnamefont {Paolo}\ \bibnamefont {Creminelli}}, \ and\ \bibinfo {author} {\bibfnamefont {Lisa}\ \bibnamefont {Randall}},\ }\bibfield  {title} {\enquote {\bibinfo {title} {{Pseudonatural inflation}},}\ }\href {\doibase 10.1088/1475-7516/2003/07/003} {\bibfield  {journal} {\bibinfo  {journal} {JCAP}\ }\textbf {\bibinfo {volume} {07}},\ \bibinfo {pages} {003} (\bibinfo {year} {2003})},\ \Eprint {http://arxiv.org/abs/hep-th/0302034} {arXiv:hep-th/0302034} \BibitemShut {NoStop}%
\bibitem [{\citenamefont {Baumann}\ and\ \citenamefont {McAllister}(2015)}]{Baumann:2014nda}%
  \BibitemOpen
  \bibfield  {author} {\bibinfo {author} {\bibfnamefont {Daniel}\ \bibnamefont {Baumann}}\ and\ \bibinfo {author} {\bibfnamefont {Liam}\ \bibnamefont {McAllister}},\ }\href {\doibase 10.1017/CBO9781316105733} {\emph {\bibinfo {title} {{Inflation and String Theory}}}},\ Cambridge Monographs on Mathematical Physics\ (\bibinfo  {publisher} {Cambridge University Press},\ \bibinfo {year} {2015})\ \Eprint {http://arxiv.org/abs/1404.2601} {arXiv:1404.2601 [hep-th]} \BibitemShut {NoStop}%
\bibitem [{\citenamefont {Bastero-Gil}\ \emph {et~al.}(2014)\citenamefont {Bastero-Gil}, \citenamefont {Berera}, \citenamefont {Moss},\ and\ \citenamefont {Ramos}}]{Bastero-Gil:2014raa}%
  \BibitemOpen
  \bibfield  {author} {\bibinfo {author} {\bibfnamefont {Mar}\ \bibnamefont {Bastero-Gil}}, \bibinfo {author} {\bibfnamefont {Arjun}\ \bibnamefont {Berera}}, \bibinfo {author} {\bibfnamefont {Ian~G.}\ \bibnamefont {Moss}}, \ and\ \bibinfo {author} {\bibfnamefont {Rudnei~O.}\ \bibnamefont {Ramos}},\ }\bibfield  {title} {\enquote {\bibinfo {title} {{Theory of non-Gaussianity in warm inflation}},}\ }\href {\doibase 10.1088/1475-7516/2014/12/008} {\bibfield  {journal} {\bibinfo  {journal} {JCAP}\ }\textbf {\bibinfo {volume} {12}},\ \bibinfo {pages} {008} (\bibinfo {year} {2014})},\ \Eprint {http://arxiv.org/abs/1408.4391} {arXiv:1408.4391 [astro-ph.CO]} \BibitemShut {NoStop}%
\bibitem [{\citenamefont {Mirbabayi}\ and\ \citenamefont {Gruzinov}(2023)}]{Mirbabayi:2022cbt}%
  \BibitemOpen
  \bibfield  {author} {\bibinfo {author} {\bibfnamefont {Mehrdad}\ \bibnamefont {Mirbabayi}}\ and\ \bibinfo {author} {\bibfnamefont {Andrei}\ \bibnamefont {Gruzinov}},\ }\bibfield  {title} {\enquote {\bibinfo {title} {{Shapes of non-Gaussianity in warm inflation}},}\ }\href {\doibase 10.1088/1475-7516/2023/02/012} {\bibfield  {journal} {\bibinfo  {journal} {JCAP}\ }\textbf {\bibinfo {volume} {02}},\ \bibinfo {pages} {012} (\bibinfo {year} {2023})},\ \Eprint {http://arxiv.org/abs/2205.13227} {arXiv:2205.13227 [astro-ph.CO]} \BibitemShut {NoStop}%
\bibitem [{\citenamefont {Yokoyama}\ and\ \citenamefont {Linde}(1999)}]{Yokoyama:1998ju}%
  \BibitemOpen
  \bibfield  {author} {\bibinfo {author} {\bibfnamefont {Junichi}\ \bibnamefont {Yokoyama}}\ and\ \bibinfo {author} {\bibfnamefont {Andrei~D.}\ \bibnamefont {Linde}},\ }\bibfield  {title} {\enquote {\bibinfo {title} {{Is warm inflation possible?}}}\ }\href {\doibase 10.1103/PhysRevD.60.083509} {\bibfield  {journal} {\bibinfo  {journal} {Phys. Rev. D}\ }\textbf {\bibinfo {volume} {60}},\ \bibinfo {pages} {083509} (\bibinfo {year} {1999})},\ \Eprint {http://arxiv.org/abs/hep-ph/9809409} {arXiv:hep-ph/9809409} \BibitemShut {NoStop}%
\bibitem [{\citenamefont {McLerran}\ \emph {et~al.}(1991)\citenamefont {McLerran}, \citenamefont {Mottola},\ and\ \citenamefont {Shaposhnikov}}]{McLerran:1990de}%
  \BibitemOpen
  \bibfield  {author} {\bibinfo {author} {\bibfnamefont {Larry~D.}\ \bibnamefont {McLerran}}, \bibinfo {author} {\bibfnamefont {Emil}\ \bibnamefont {Mottola}}, \ and\ \bibinfo {author} {\bibfnamefont {Mikhail~E.}\ \bibnamefont {Shaposhnikov}},\ }\bibfield  {title} {\enquote {\bibinfo {title} {{Sphalerons and Axion Dynamics in High Temperature {QCD}}},}\ }\href {\doibase 10.1103/PhysRevD.43.2027} {\bibfield  {journal} {\bibinfo  {journal} {Phys. Rev. D}\ }\textbf {\bibinfo {volume} {43}},\ \bibinfo {pages} {2027--2035} (\bibinfo {year} {1991})}\BibitemShut {NoStop}%
\bibitem [{\citenamefont {Berghaus}\ \emph {et~al.}(2020)\citenamefont {Berghaus}, \citenamefont {Graham},\ and\ \citenamefont {Kaplan}}]{Berghaus:2019whh}%
  \BibitemOpen
  \bibfield  {author} {\bibinfo {author} {\bibfnamefont {Kim~V.}\ \bibnamefont {Berghaus}}, \bibinfo {author} {\bibfnamefont {Peter~W.}\ \bibnamefont {Graham}}, \ and\ \bibinfo {author} {\bibfnamefont {David~E.}\ \bibnamefont {Kaplan}},\ }\bibfield  {title} {\enquote {\bibinfo {title} {{Minimal Warm Inflation}},}\ }\href {\doibase 10.1088/1475-7516/2020/03/034} {\bibfield  {journal} {\bibinfo  {journal} {JCAP}\ }\textbf {\bibinfo {volume} {03}},\ \bibinfo {pages} {034} (\bibinfo {year} {2020})},\ \bibinfo {note} {[Erratum: JCAP 10, E02 (2023)]},\ \Eprint {http://arxiv.org/abs/1910.07525} {arXiv:1910.07525 [hep-ph]} \BibitemShut {NoStop}%
\bibitem [{\citenamefont {Laine}\ and\ \citenamefont {Procacci}(2021)}]{Laine:2021ego}%
  \BibitemOpen
  \bibfield  {author} {\bibinfo {author} {\bibfnamefont {M.}~\bibnamefont {Laine}}\ and\ \bibinfo {author} {\bibfnamefont {S.}~\bibnamefont {Procacci}},\ }\bibfield  {title} {\enquote {\bibinfo {title} {{Minimal warm inflation with complete medium response}},}\ }\href {\doibase 10.1088/1475-7516/2021/06/031} {\bibfield  {journal} {\bibinfo  {journal} {JCAP}\ }\textbf {\bibinfo {volume} {06}},\ \bibinfo {pages} {031} (\bibinfo {year} {2021})},\ \Eprint {http://arxiv.org/abs/2102.09913} {arXiv:2102.09913 [hep-ph]} \BibitemShut {NoStop}%
\bibitem [{\citenamefont {Berghaus}\ \emph {et~al.}(2021)\citenamefont {Berghaus}, \citenamefont {Graham}, \citenamefont {Kaplan}, \citenamefont {Moore},\ and\ \citenamefont {Rajendran}}]{Berghaus:2020ekh}%
  \BibitemOpen
  \bibfield  {author} {\bibinfo {author} {\bibfnamefont {Kim~V.}\ \bibnamefont {Berghaus}}, \bibinfo {author} {\bibfnamefont {Peter~W.}\ \bibnamefont {Graham}}, \bibinfo {author} {\bibfnamefont {David~E.}\ \bibnamefont {Kaplan}}, \bibinfo {author} {\bibfnamefont {Guy~D.}\ \bibnamefont {Moore}}, \ and\ \bibinfo {author} {\bibfnamefont {Surjeet}\ \bibnamefont {Rajendran}},\ }\bibfield  {title} {\enquote {\bibinfo {title} {{Dark energy radiation}},}\ }\href {\doibase 10.1103/PhysRevD.104.083520} {\bibfield  {journal} {\bibinfo  {journal} {Phys. Rev. D}\ }\textbf {\bibinfo {volume} {104}},\ \bibinfo {pages} {083520} (\bibinfo {year} {2021})},\ \Eprint {http://arxiv.org/abs/2012.10549} {arXiv:2012.10549 [hep-ph]} \BibitemShut {NoStop}%
\bibitem [{\citenamefont {Berghaus}\ \emph {et~al.}(2024)\citenamefont {Berghaus}, \citenamefont {Forslund},\ and\ \citenamefont {Guevarra}}]{Berghaus:2024zfg}%
  \BibitemOpen
  \bibfield  {author} {\bibinfo {author} {\bibfnamefont {Kim~V.}\ \bibnamefont {Berghaus}}, \bibinfo {author} {\bibfnamefont {Matthew}\ \bibnamefont {Forslund}}, \ and\ \bibinfo {author} {\bibfnamefont {Mark~Vincent}\ \bibnamefont {Guevarra}},\ }\bibfield  {title} {\enquote {\bibinfo {title} {{Warm inflation with a heavy QCD axion}},}\ }\href {\doibase 10.1088/1475-7516/2024/10/103} {\bibfield  {journal} {\bibinfo  {journal} {JCAP}\ }\textbf {\bibinfo {volume} {10}},\ \bibinfo {pages} {103} (\bibinfo {year} {2024})},\ \Eprint {http://arxiv.org/abs/2402.13535} {arXiv:2402.13535 [hep-ph]} \BibitemShut {NoStop}%
\bibitem [{\citenamefont {Drewes}\ and\ \citenamefont {Zell}(2024)}]{Drewes:2023khq}%
  \BibitemOpen
  \bibfield  {author} {\bibinfo {author} {\bibfnamefont {Marco}\ \bibnamefont {Drewes}}\ and\ \bibinfo {author} {\bibfnamefont {Sebastian}\ \bibnamefont {Zell}},\ }\bibfield  {title} {\enquote {\bibinfo {title} {{On sphaleron heating in the presence of fermions}},}\ }\href {\doibase 10.1088/1475-7516/2024/06/038} {\bibfield  {journal} {\bibinfo  {journal} {JCAP}\ }\textbf {\bibinfo {volume} {06}},\ \bibinfo {pages} {038} (\bibinfo {year} {2024})},\ \Eprint {http://arxiv.org/abs/2312.13739} {arXiv:2312.13739 [hep-ph]} \BibitemShut {NoStop}%
\bibitem [{\citenamefont {Buldgen}\ \emph {et~al.}(2022)\citenamefont {Buldgen}, \citenamefont {Drewes}, \citenamefont {Kang},\ and\ \citenamefont {Mun}}]{Buldgen:2019dus}%
  \BibitemOpen
  \bibfield  {author} {\bibinfo {author} {\bibfnamefont {Gilles}\ \bibnamefont {Buldgen}}, \bibinfo {author} {\bibfnamefont {Marco}\ \bibnamefont {Drewes}}, \bibinfo {author} {\bibfnamefont {Jin~U.}\ \bibnamefont {Kang}}, \ and\ \bibinfo {author} {\bibfnamefont {Ui~Ri}\ \bibnamefont {Mun}},\ }\bibfield  {title} {\enquote {\bibinfo {title} {{General Markovian equation for scalar fields in a slowly evolving background}},}\ }\href {\doibase 10.1088/1475-7516/2022/05/039} {\bibfield  {journal} {\bibinfo  {journal} {JCAP}\ }\textbf {\bibinfo {volume} {05}},\ \bibinfo {pages} {039} (\bibinfo {year} {2022})},\ \Eprint {http://arxiv.org/abs/1912.02772} {arXiv:1912.02772 [hep-ph]} \BibitemShut {NoStop}%
\bibitem [{\citenamefont {Berera}\ \emph {et~al.}(2009)\citenamefont {Berera}, \citenamefont {Moss},\ and\ \citenamefont {Ramos}}]{Berera:2008ar}%
  \BibitemOpen
  \bibfield  {author} {\bibinfo {author} {\bibfnamefont {Arjun}\ \bibnamefont {Berera}}, \bibinfo {author} {\bibfnamefont {Ian~G.}\ \bibnamefont {Moss}}, \ and\ \bibinfo {author} {\bibfnamefont {Rudnei~O.}\ \bibnamefont {Ramos}},\ }\bibfield  {title} {\enquote {\bibinfo {title} {{Warm Inflation and its Microphysical Basis}},}\ }\href {\doibase 10.1088/0034-4885/72/2/026901} {\bibfield  {journal} {\bibinfo  {journal} {Rept. Prog. Phys.}\ }\textbf {\bibinfo {volume} {72}},\ \bibinfo {pages} {026901} (\bibinfo {year} {2009})},\ \Eprint {http://arxiv.org/abs/0808.1855} {arXiv:0808.1855 [hep-ph]} \BibitemShut {NoStop}%
\bibitem [{\citenamefont {Laine}\ and\ \citenamefont {Vuorinen}(2016)}]{Laine:2016hma}%
  \BibitemOpen
  \bibfield  {author} {\bibinfo {author} {\bibfnamefont {Mikko}\ \bibnamefont {Laine}}\ and\ \bibinfo {author} {\bibfnamefont {Aleksi}\ \bibnamefont {Vuorinen}},\ }\href {\doibase 10.1007/978-3-319-31933-9} {\emph {\bibinfo {title} {{Basics of Thermal Field Theory}}}},\ Vol.\ \bibinfo {volume} {925}\ (\bibinfo  {publisher} {Springer},\ \bibinfo {year} {2016})\ \Eprint {http://arxiv.org/abs/1701.01554} {arXiv:1701.01554 [hep-ph]} \BibitemShut {NoStop}%
\bibitem [{\citenamefont {Moore}\ and\ \citenamefont {Tassler}(2011)}]{Moore:2010jd}%
  \BibitemOpen
  \bibfield  {author} {\bibinfo {author} {\bibfnamefont {Guy~D.}\ \bibnamefont {Moore}}\ and\ \bibinfo {author} {\bibfnamefont {Marcus}\ \bibnamefont {Tassler}},\ }\bibfield  {title} {\enquote {\bibinfo {title} {{The Sphaleron Rate in SU(N) Gauge Theory}},}\ }\href {\doibase 10.1007/JHEP02(2011)105} {\bibfield  {journal} {\bibinfo  {journal} {JHEP}\ }\textbf {\bibinfo {volume} {02}},\ \bibinfo {pages} {105} (\bibinfo {year} {2011})},\ \Eprint {http://arxiv.org/abs/1011.1167} {arXiv:1011.1167 [hep-ph]} \BibitemShut {NoStop}%
\bibitem [{\citenamefont {Laine}\ \emph {et~al.}(2022)\citenamefont {Laine}, \citenamefont {Niemi}, \citenamefont {Procacci},\ and\ \citenamefont {Rummukainen}}]{Laine:2022ytc}%
  \BibitemOpen
  \bibfield  {author} {\bibinfo {author} {\bibfnamefont {M.}~\bibnamefont {Laine}}, \bibinfo {author} {\bibfnamefont {L.}~\bibnamefont {Niemi}}, \bibinfo {author} {\bibfnamefont {S.}~\bibnamefont {Procacci}}, \ and\ \bibinfo {author} {\bibfnamefont {K.}~\bibnamefont {Rummukainen}},\ }\bibfield  {title} {\enquote {\bibinfo {title} {{Shape of the hot topological charge density spectral function}},}\ }\href {\doibase 10.1007/JHEP11(2022)126} {\bibfield  {journal} {\bibinfo  {journal} {JHEP}\ }\textbf {\bibinfo {volume} {11}},\ \bibinfo {pages} {126} (\bibinfo {year} {2022})},\ \Eprint {http://arxiv.org/abs/2209.13804} {arXiv:2209.13804 [hep-ph]} \BibitemShut {NoStop}%
\bibitem [{\citenamefont {Klose}\ \emph {et~al.}(2022)\citenamefont {Klose}, \citenamefont {Laine},\ and\ \citenamefont {Procacci}}]{Klose:2022rxh}%
  \BibitemOpen
  \bibfield  {author} {\bibinfo {author} {\bibfnamefont {P.}~\bibnamefont {Klose}}, \bibinfo {author} {\bibfnamefont {M.}~\bibnamefont {Laine}}, \ and\ \bibinfo {author} {\bibfnamefont {S.}~\bibnamefont {Procacci}},\ }\bibfield  {title} {\enquote {\bibinfo {title} {{Gravitational wave background from vacuum and thermal fluctuations during axion-like inflation}},}\ }\href {\doibase 10.1088/1475-7516/2022/12/020} {\bibfield  {journal} {\bibinfo  {journal} {JCAP}\ }\textbf {\bibinfo {volume} {12}},\ \bibinfo {pages} {020} (\bibinfo {year} {2022})},\ \Eprint {http://arxiv.org/abs/2210.11710} {arXiv:2210.11710 [hep-ph]} \BibitemShut {NoStop}%
\bibitem [{\citenamefont {Kolesova}\ \emph {et~al.}(2023)\citenamefont {Kolesova}, \citenamefont {Laine},\ and\ \citenamefont {Procacci}}]{Kolesova:2023yfp}%
  \BibitemOpen
  \bibfield  {author} {\bibinfo {author} {\bibfnamefont {H.}~\bibnamefont {Kolesova}}, \bibinfo {author} {\bibfnamefont {M.}~\bibnamefont {Laine}}, \ and\ \bibinfo {author} {\bibfnamefont {S.}~\bibnamefont {Procacci}},\ }\bibfield  {title} {\enquote {\bibinfo {title} {{Maximal temperature of strongly-coupled dark sectors}},}\ }\href {\doibase 10.1007/JHEP05(2023)239} {\bibfield  {journal} {\bibinfo  {journal} {JHEP}\ }\textbf {\bibinfo {volume} {05}},\ \bibinfo {pages} {239} (\bibinfo {year} {2023})},\ \Eprint {http://arxiv.org/abs/2303.17973} {arXiv:2303.17973 [hep-ph]} \BibitemShut {NoStop}%
\bibitem [{\citenamefont {Coleman}(1985)}]{Coleman:1985rnk}%
  \BibitemOpen
  \bibfield  {author} {\bibinfo {author} {\bibfnamefont {Sidney}\ \bibnamefont {Coleman}},\ }\href {\doibase 10.1017/CBO9780511565045} {\emph {\bibinfo {title} {{Aspects of Symmetry}: {Selected Erice Lectures}}}}\ (\bibinfo  {publisher} {Cambridge University Press},\ \bibinfo {address} {Cambridge, U.K.},\ \bibinfo {year} {1985})\BibitemShut {NoStop}%
\bibitem [{\citenamefont {Zell}(2025)}]{Zell:2024vfn}%
  \BibitemOpen
  \bibfield  {author} {\bibinfo {author} {\bibfnamefont {Sebastian}\ \bibnamefont {Zell}},\ }\bibfield  {title} {\enquote {\bibinfo {title} {{No Warm Inflation From a Vanilla Axion}},}\ }\href {\doibase 10.1103/kp15-3s1c} {\bibfield  {journal} {\bibinfo  {journal} {Phys. Rev. D}\ }\textbf {\bibinfo {volume} {112}},\ \bibinfo {pages} {L081307} (\bibinfo {year} {2025})},\ \Eprint {http://arxiv.org/abs/2408.07746} {arXiv:2408.07746 [hep-ph]} \BibitemShut {NoStop}%
\bibitem [{\citenamefont {Peskin}\ and\ \citenamefont {Schroeder}(1995)}]{Peskin:1995ev}%
  \BibitemOpen
  \bibfield  {author} {\bibinfo {author} {\bibfnamefont {Michael~E.}\ \bibnamefont {Peskin}}\ and\ \bibinfo {author} {\bibfnamefont {Daniel~V.}\ \bibnamefont {Schroeder}},\ }\href {\doibase 10.1201/9780429503559} {\emph {\bibinfo {title} {{An Introduction to quantum field theory}}}}\ (\bibinfo  {publisher} {Addison-Wesley},\ \bibinfo {address} {Reading, USA},\ \bibinfo {year} {1995})\BibitemShut {NoStop}%
\bibitem [{\citenamefont {Khlebnikov}\ and\ \citenamefont {Shaposhnikov}(1988)}]{Khlebnikov:1988sr}%
  \BibitemOpen
  \bibfield  {author} {\bibinfo {author} {\bibfnamefont {S.~Yu.}\ \bibnamefont {Khlebnikov}}\ and\ \bibinfo {author} {\bibfnamefont {M.~E.}\ \bibnamefont {Shaposhnikov}},\ }\bibfield  {title} {\enquote {\bibinfo {title} {{The Statistical Theory of Anomalous Fermion Number Nonconservation}},}\ }\href {\doibase 10.1016/0550-3213(88)90133-2} {\bibfield  {journal} {\bibinfo  {journal} {Nucl. Phys. B}\ }\textbf {\bibinfo {volume} {308}},\ \bibinfo {pages} {885--912} (\bibinfo {year} {1988})}\BibitemShut {NoStop}%
\bibitem [{\citenamefont {Giudice}\ and\ \citenamefont {Shaposhnikov}(1994)}]{Giudice:1993bb}%
  \BibitemOpen
  \bibfield  {author} {\bibinfo {author} {\bibfnamefont {G.~F.}\ \bibnamefont {Giudice}}\ and\ \bibinfo {author} {\bibfnamefont {Mikhail~E.}\ \bibnamefont {Shaposhnikov}},\ }\bibfield  {title} {\enquote {\bibinfo {title} {{Strong sphalerons and electroweak baryogenesis}},}\ }\href {\doibase 10.1016/0370-2693(94)91202-5} {\bibfield  {journal} {\bibinfo  {journal} {Phys. Lett. B}\ }\textbf {\bibinfo {volume} {326}},\ \bibinfo {pages} {118--124} (\bibinfo {year} {1994})},\ \Eprint {http://arxiv.org/abs/hep-ph/9311367} {arXiv:hep-ph/9311367} \BibitemShut {NoStop}%
\bibitem [{\citenamefont {Rubakov}\ and\ \citenamefont {Shaposhnikov}(1996)}]{Rubakov:1996vz}%
  \BibitemOpen
  \bibfield  {author} {\bibinfo {author} {\bibfnamefont {V.~A.}\ \bibnamefont {Rubakov}}\ and\ \bibinfo {author} {\bibfnamefont {M.~E.}\ \bibnamefont {Shaposhnikov}},\ }\bibfield  {title} {\enquote {\bibinfo {title} {{Electroweak baryon number nonconservation in the early universe and in high-energy collisions}},}\ }\href {\doibase 10.1070/PU1996v039n05ABEH000145} {\bibfield  {journal} {\bibinfo  {journal} {Usp. Fiz. Nauk}\ }\textbf {\bibinfo {volume} {166}},\ \bibinfo {pages} {493--537} (\bibinfo {year} {1996})},\ \Eprint {http://arxiv.org/abs/hep-ph/9603208} {arXiv:hep-ph/9603208} \BibitemShut {NoStop}%
\bibitem [{\citenamefont {Moore}(1996)}]{Moore:1996qs}%
  \BibitemOpen
  \bibfield  {author} {\bibinfo {author} {\bibfnamefont {Guy~D.}\ \bibnamefont {Moore}},\ }\bibfield  {title} {\enquote {\bibinfo {title} {{Motion of Chern-Simons number at high temperatures under a chemical potential}},}\ }\href {\doibase 10.1016/S0550-3213(96)00445-2} {\bibfield  {journal} {\bibinfo  {journal} {Nucl. Phys. B}\ }\textbf {\bibinfo {volume} {480}},\ \bibinfo {pages} {657--688} (\bibinfo {year} {1996})},\ \Eprint {http://arxiv.org/abs/hep-ph/9603384} {arXiv:hep-ph/9603384} \BibitemShut {NoStop}%
\bibitem [{\citenamefont {Moore}(1997)}]{Moore:1997im}%
  \BibitemOpen
  \bibfield  {author} {\bibinfo {author} {\bibfnamefont {Guy~D.}\ \bibnamefont {Moore}},\ }\bibfield  {title} {\enquote {\bibinfo {title} {{Computing the strong sphaleron rate}},}\ }\href {\doibase 10.1016/S0370-2693(97)01046-0} {\bibfield  {journal} {\bibinfo  {journal} {Phys. Lett. B}\ }\textbf {\bibinfo {volume} {412}},\ \bibinfo {pages} {359--370} (\bibinfo {year} {1997})},\ \Eprint {http://arxiv.org/abs/hep-ph/9705248} {arXiv:hep-ph/9705248} \BibitemShut {NoStop}%
\bibitem [{\citenamefont {Boyarsky}\ \emph {et~al.}(2021)\citenamefont {Boyarsky}, \citenamefont {Cheianov}, \citenamefont {Ruchayskiy},\ and\ \citenamefont {Sobol}}]{Boyarsky:2020cyk}%
  \BibitemOpen
  \bibfield  {author} {\bibinfo {author} {\bibfnamefont {A.}~\bibnamefont {Boyarsky}}, \bibinfo {author} {\bibfnamefont {V.}~\bibnamefont {Cheianov}}, \bibinfo {author} {\bibfnamefont {O.}~\bibnamefont {Ruchayskiy}}, \ and\ \bibinfo {author} {\bibfnamefont {O.}~\bibnamefont {Sobol}},\ }\bibfield  {title} {\enquote {\bibinfo {title} {{Evolution of the Primordial Axial Charge across Cosmic Times}},}\ }\href {\doibase 10.1103/PhysRevLett.126.021801} {\bibfield  {journal} {\bibinfo  {journal} {Phys. Rev. Lett.}\ }\textbf {\bibinfo {volume} {126}},\ \bibinfo {pages} {021801} (\bibinfo {year} {2021})},\ \Eprint {http://arxiv.org/abs/2007.13691} {arXiv:2007.13691 [hep-ph]} \BibitemShut {NoStop}%
\bibitem [{\citenamefont {B\"odeker}\ and\ \citenamefont {Schr\"oder}(2019)}]{Bodeker:2019ajh}%
  \BibitemOpen
  \bibfield  {author} {\bibinfo {author} {\bibfnamefont {Dietrich}\ \bibnamefont {B\"odeker}}\ and\ \bibinfo {author} {\bibfnamefont {Dennis}\ \bibnamefont {Schr\"oder}},\ }\bibfield  {title} {\enquote {\bibinfo {title} {{Equilibration of right-handed electrons}},}\ }\href {\doibase 10.1088/1475-7516/2019/05/010} {\bibfield  {journal} {\bibinfo  {journal} {JCAP}\ }\textbf {\bibinfo {volume} {05}},\ \bibinfo {pages} {010} (\bibinfo {year} {2019})},\ \Eprint {http://arxiv.org/abs/1902.07220} {arXiv:1902.07220 [hep-ph]} \BibitemShut {NoStop}%
\bibitem [{\citenamefont {Sartore}\ and\ \citenamefont {Schienbein}(2021)}]{Sartore:2020gou}%
  \BibitemOpen
  \bibfield  {author} {\bibinfo {author} {\bibfnamefont {Lohan}\ \bibnamefont {Sartore}}\ and\ \bibinfo {author} {\bibfnamefont {Ingo}\ \bibnamefont {Schienbein}},\ }\bibfield  {title} {\enquote {\bibinfo {title} {{PyR@TE 3}},}\ }\href {\doibase 10.1016/j.cpc.2020.107819} {\bibfield  {journal} {\bibinfo  {journal} {Comput. Phys. Commun.}\ }\textbf {\bibinfo {volume} {261}},\ \bibinfo {pages} {107819} (\bibinfo {year} {2021})},\ \Eprint {http://arxiv.org/abs/2007.12700} {arXiv:2007.12700 [hep-ph]} \BibitemShut {NoStop}%
\bibitem [{\citenamefont {Kaczmarek}\ and\ \citenamefont {Zantow}(2005)}]{Kaczmarek:2005ui}%
  \BibitemOpen
  \bibfield  {author} {\bibinfo {author} {\bibfnamefont {Olaf}\ \bibnamefont {Kaczmarek}}\ and\ \bibinfo {author} {\bibfnamefont {Felix}\ \bibnamefont {Zantow}},\ }\bibfield  {title} {\enquote {\bibinfo {title} {{Static quark anti-quark interactions in zero and finite temperature QCD. I. Heavy quark free energies, running coupling and quarkonium binding}},}\ }\href {\doibase 10.1103/PhysRevD.71.114510} {\bibfield  {journal} {\bibinfo  {journal} {Phys. Rev. D}\ }\textbf {\bibinfo {volume} {71}},\ \bibinfo {pages} {114510} (\bibinfo {year} {2005})},\ \Eprint {http://arxiv.org/abs/hep-lat/0503017} {arXiv:hep-lat/0503017} \BibitemShut {NoStop}%
\bibitem [{\citenamefont {Braun}\ and\ \citenamefont {Gies}(2007)}]{Braun:2005uj}%
  \BibitemOpen
  \bibfield  {author} {\bibinfo {author} {\bibfnamefont {Jens}\ \bibnamefont {Braun}}\ and\ \bibinfo {author} {\bibfnamefont {Holger}\ \bibnamefont {Gies}},\ }\bibfield  {title} {\enquote {\bibinfo {title} {{Running coupling at finite temperature and chiral symmetry restoration in QCD}},}\ }\href {\doibase 10.1016/j.physletb.2006.11.059} {\bibfield  {journal} {\bibinfo  {journal} {Phys. Lett. B}\ }\textbf {\bibinfo {volume} {645}},\ \bibinfo {pages} {53--58} (\bibinfo {year} {2007})},\ \Eprint {http://arxiv.org/abs/hep-ph/0512085} {arXiv:hep-ph/0512085} \BibitemShut {NoStop}%
\bibitem [{\citenamefont {Li}\ \emph {et~al.}(2019)\citenamefont {Li} \emph {et~al.}}]{Li:2017drr}%
  \BibitemOpen
  \bibfield  {author} {\bibinfo {author} {\bibfnamefont {Hong}\ \bibnamefont {Li}} \emph {et~al.},\ }\bibfield  {title} {\enquote {\bibinfo {title} {{Probing Primordial Gravitational Waves: Ali CMB Polarization Telescope}},}\ }\href {\doibase 10.1093/nsr/nwy019} {\bibfield  {journal} {\bibinfo  {journal} {Natl. Sci. Rev.}\ }\textbf {\bibinfo {volume} {6}},\ \bibinfo {pages} {145--154} (\bibinfo {year} {2019})},\ \Eprint {http://arxiv.org/abs/1710.03047} {arXiv:1710.03047 [astro-ph.CO]} \BibitemShut {NoStop}%
\bibitem [{\citenamefont {Liu}\ \emph {et~al.}(2025)\citenamefont {Liu}, \citenamefont {Ming}, \citenamefont {Drewes},\ and\ \citenamefont {Li}}]{Liu:2025sut}%
  \BibitemOpen
  \bibfield  {author} {\bibinfo {author} {\bibfnamefont {Yang}\ \bibnamefont {Liu}}, \bibinfo {author} {\bibfnamefont {Lei}\ \bibnamefont {Ming}}, \bibinfo {author} {\bibfnamefont {Marco}\ \bibnamefont {Drewes}}, \ and\ \bibinfo {author} {\bibfnamefont {Hong}\ \bibnamefont {Li}},\ }\bibfield  {title} {\enquote {\bibinfo {title} {{AliCPT Sensitivity to Cosmic Reheating}},}\ }\href@noop {} {\  (\bibinfo {year} {2025})},\ \Eprint {http://arxiv.org/abs/2503.21207} {arXiv:2503.21207 [astro-ph.CO]} \BibitemShut {NoStop}%
\bibitem [{\citenamefont {Ade}\ \emph {et~al.}(2019)\citenamefont {Ade} \emph {et~al.}}]{SimonsObservatory:2018koc}%
  \BibitemOpen
  \bibfield  {author} {\bibinfo {author} {\bibfnamefont {Peter}\ \bibnamefont {Ade}} \emph {et~al.} (\bibinfo {collaboration} {Simons Observatory}),\ }\bibfield  {title} {\enquote {\bibinfo {title} {{The Simons Observatory: Science goals and forecasts}},}\ }\href {\doibase 10.1088/1475-7516/2019/02/056} {\bibfield  {journal} {\bibinfo  {journal} {JCAP}\ }\textbf {\bibinfo {volume} {02}},\ \bibinfo {pages} {056} (\bibinfo {year} {2019})},\ \Eprint {http://arxiv.org/abs/1808.07445} {arXiv:1808.07445 [astro-ph.CO]} \BibitemShut {NoStop}%
\bibitem [{\citenamefont {Suzuki}\ \emph {et~al.}(2016)\citenamefont {Suzuki} \emph {et~al.}}]{POLARBEAR:2015ixw}%
  \BibitemOpen
  \bibfield  {author} {\bibinfo {author} {\bibfnamefont {A.}~\bibnamefont {Suzuki}} \emph {et~al.} (\bibinfo {collaboration} {POLARBEAR}),\ }\bibfield  {title} {\enquote {\bibinfo {title} {{The POLARBEAR-2 and the Simons Array Experiment}},}\ }\href {\doibase 10.1007/s10909-015-1425-4} {\bibfield  {journal} {\bibinfo  {journal} {J. Low Temp. Phys.}\ }\textbf {\bibinfo {volume} {184}},\ \bibinfo {pages} {805--810} (\bibinfo {year} {2016})},\ \Eprint {http://arxiv.org/abs/1512.07299} {arXiv:1512.07299 [astro-ph.IM]} \BibitemShut {NoStop}%
\bibitem [{\citenamefont {Allys}\ \emph {et~al.}(2023)\citenamefont {Allys} \emph {et~al.}}]{LiteBIRD:2022cnt}%
  \BibitemOpen
  \bibfield  {author} {\bibinfo {author} {\bibfnamefont {E.}~\bibnamefont {Allys}} \emph {et~al.} (\bibinfo {collaboration} {LiteBIRD}),\ }\bibfield  {title} {\enquote {\bibinfo {title} {{Probing Cosmic Inflation with the LiteBIRD Cosmic Microwave Background Polarization Survey}},}\ }\href {\doibase 10.1093/ptep/ptac150} {\bibfield  {journal} {\bibinfo  {journal} {PTEP}\ }\textbf {\bibinfo {volume} {2023}},\ \bibinfo {pages} {042F01} (\bibinfo {year} {2023})},\ \Eprint {http://arxiv.org/abs/2202.02773} {arXiv:2202.02773 [astro-ph.IM]} \BibitemShut {NoStop}%
\bibitem [{\citenamefont {Abazajian}\ \emph {et~al.}(2022)\citenamefont {Abazajian} \emph {et~al.}}]{CMB-S4:2020lpa}%
  \BibitemOpen
  \bibfield  {author} {\bibinfo {author} {\bibfnamefont {Kevork}\ \bibnamefont {Abazajian}} \emph {et~al.} (\bibinfo {collaboration} {CMB-S4}),\ }\bibfield  {title} {\enquote {\bibinfo {title} {{CMB-S4: Forecasting Constraints on Primordial Gravitational Waves}},}\ }\href {\doibase 10.3847/1538-4357/ac1596} {\bibfield  {journal} {\bibinfo  {journal} {Astrophys. J.}\ }\textbf {\bibinfo {volume} {926}},\ \bibinfo {pages} {54} (\bibinfo {year} {2022})},\ \Eprint {http://arxiv.org/abs/2008.12619} {arXiv:2008.12619 [astro-ph.CO]} \BibitemShut {NoStop}%
\bibitem [{\citenamefont {Drewes}(2022)}]{Drewes:2019rxn}%
  \BibitemOpen
  \bibfield  {author} {\bibinfo {author} {\bibfnamefont {Marco}\ \bibnamefont {Drewes}},\ }\bibfield  {title} {\enquote {\bibinfo {title} {{Measuring the inflaton coupling in~the~CMB}},}\ }\href {\doibase 10.1088/1475-7516/2022/09/069} {\bibfield  {journal} {\bibinfo  {journal} {JCAP}\ }\textbf {\bibinfo {volume} {09}},\ \bibinfo {pages} {069} (\bibinfo {year} {2022})},\ \Eprint {http://arxiv.org/abs/1903.09599} {arXiv:1903.09599 [astro-ph.CO]} \BibitemShut {NoStop}%
\bibitem [{\citenamefont {Drewes}\ and\ \citenamefont {Ming}(2024)}]{Drewes:2022nhu}%
  \BibitemOpen
  \bibfield  {author} {\bibinfo {author} {\bibfnamefont {Marco}\ \bibnamefont {Drewes}}\ and\ \bibinfo {author} {\bibfnamefont {Lei}\ \bibnamefont {Ming}},\ }\bibfield  {title} {\enquote {\bibinfo {title} {{Connecting Cosmic Inflation to Particle Physics with LiteBIRD, CMB-S4, EUCLID, and SKA}},}\ }\href {\doibase 10.1103/PhysRevLett.133.031001} {\bibfield  {journal} {\bibinfo  {journal} {Phys. Rev. Lett.}\ }\textbf {\bibinfo {volume} {133}},\ \bibinfo {pages} {031001} (\bibinfo {year} {2024})},\ \Eprint {http://arxiv.org/abs/2208.07609} {arXiv:2208.07609 [hep-ph]} \BibitemShut {NoStop}%
\bibitem [{\citenamefont {Bezrukov}\ and\ \citenamefont {Shaposhnikov}(2008)}]{Bezrukov:2007ep}%
  \BibitemOpen
  \bibfield  {author} {\bibinfo {author} {\bibfnamefont {Fedor~L.}\ \bibnamefont {Bezrukov}}\ and\ \bibinfo {author} {\bibfnamefont {Mikhail}\ \bibnamefont {Shaposhnikov}},\ }\bibfield  {title} {\enquote {\bibinfo {title} {{The Standard Model Higgs boson as the inflaton}},}\ }\href {\doibase 10.1016/j.physletb.2007.11.072} {\bibfield  {journal} {\bibinfo  {journal} {Phys. Lett. B}\ }\textbf {\bibinfo {volume} {659}},\ \bibinfo {pages} {703--706} (\bibinfo {year} {2008})},\ \Eprint {http://arxiv.org/abs/0710.3755} {arXiv:0710.3755 [hep-th]} \BibitemShut {NoStop}%
\bibitem [{\citenamefont {Shaposhnikov}\ \emph {et~al.}(2020)\citenamefont {Shaposhnikov}, \citenamefont {Shkerin},\ and\ \citenamefont {Zell}}]{Shaposhnikov:2020fdv}%
  \BibitemOpen
  \bibfield  {author} {\bibinfo {author} {\bibfnamefont {Mikhail}\ \bibnamefont {Shaposhnikov}}, \bibinfo {author} {\bibfnamefont {Andrey}\ \bibnamefont {Shkerin}}, \ and\ \bibinfo {author} {\bibfnamefont {Sebastian}\ \bibnamefont {Zell}},\ }\bibfield  {title} {\enquote {\bibinfo {title} {{Quantum Effects in Palatini Higgs Inflation}},}\ }\href {\doibase 10.1088/1475-7516/2020/07/064} {\bibfield  {journal} {\bibinfo  {journal} {JCAP}\ }\textbf {\bibinfo {volume} {07}},\ \bibinfo {pages} {064} (\bibinfo {year} {2020})},\ \Eprint {http://arxiv.org/abs/2002.07105} {arXiv:2002.07105 [hep-ph]} \BibitemShut {NoStop}%
\bibitem [{\citenamefont {Antel}\ \emph {et~al.}(2023)\citenamefont {Antel} \emph {et~al.}}]{Antel:2023hkf}%
  \BibitemOpen
  \bibfield  {author} {\bibinfo {author} {\bibfnamefont {C.}~\bibnamefont {Antel}} \emph {et~al.},\ }\bibfield  {title} {\enquote {\bibinfo {title} {{Feebly-interacting particles: FIPs 2022 Workshop Report}},}\ }\href {\doibase 10.1140/epjc/s10052-023-12168-5} {\bibfield  {journal} {\bibinfo  {journal} {Eur. Phys. J. C}\ }\textbf {\bibinfo {volume} {83}},\ \bibinfo {pages} {1122} (\bibinfo {year} {2023})},\ \Eprint {http://arxiv.org/abs/2305.01715} {arXiv:2305.01715 [hep-ph]} \BibitemShut {NoStop}%
\bibitem [{\citenamefont {O'Hare}(2020)}]{AxionLimits}%
  \BibitemOpen
  \bibfield  {author} {\bibinfo {author} {\bibfnamefont {Ciaran}\ \bibnamefont {O'Hare}},\ }\href {\doibase 10.5281/zenodo.3932430} {\enquote {\bibinfo {title} {cajohare/axionlimits: Axionlimits},}\ }\bibinfo {howpublished} {\url{https://cajohare.github.io/AxionLimits/}} (\bibinfo {year} {2020})\BibitemShut {NoStop}%
\bibitem [{\citenamefont {Peccei}\ and\ \citenamefont {Quinn}(1977)}]{Peccei:1977hh}%
  \BibitemOpen
  \bibfield  {author} {\bibinfo {author} {\bibfnamefont {R.~D.}\ \bibnamefont {Peccei}}\ and\ \bibinfo {author} {\bibfnamefont {Helen~R.}\ \bibnamefont {Quinn}},\ }\bibfield  {title} {\enquote {\bibinfo {title} {{CP Conservation in the Presence of Instantons}},}\ }\href {\doibase 10.1103/PhysRevLett.38.1440} {\bibfield  {journal} {\bibinfo  {journal} {Phys. Rev. Lett.}\ }\textbf {\bibinfo {volume} {38}},\ \bibinfo {pages} {1440--1443} (\bibinfo {year} {1977})}\BibitemShut {NoStop}%
\bibitem [{\citenamefont {Weinberg}(1978)}]{Weinberg:1977ma}%
  \BibitemOpen
  \bibfield  {author} {\bibinfo {author} {\bibfnamefont {Steven}\ \bibnamefont {Weinberg}},\ }\bibfield  {title} {\enquote {\bibinfo {title} {{A New Light Boson?}}}\ }\href {\doibase 10.1103/PhysRevLett.40.223} {\bibfield  {journal} {\bibinfo  {journal} {Phys. Rev. Lett.}\ }\textbf {\bibinfo {volume} {40}},\ \bibinfo {pages} {223--226} (\bibinfo {year} {1978})}\BibitemShut {NoStop}%
\bibitem [{\citenamefont {Wilczek}(1978)}]{Wilczek:1977pj}%
  \BibitemOpen
  \bibfield  {author} {\bibinfo {author} {\bibfnamefont {Frank}\ \bibnamefont {Wilczek}},\ }\bibfield  {title} {\enquote {\bibinfo {title} {{Problem of Strong $P$ and $T$ Invariance in the Presence of Instantons}},}\ }\href {\doibase 10.1103/PhysRevLett.40.279} {\bibfield  {journal} {\bibinfo  {journal} {Phys. Rev. Lett.}\ }\textbf {\bibinfo {volume} {40}},\ \bibinfo {pages} {279--282} (\bibinfo {year} {1978})}\BibitemShut {NoStop}%
\bibitem [{\citenamefont {O.~Ramos}\ and\ \citenamefont {Rodrigues}(2025)}]{ORamos:2025uqs}%
  \BibitemOpen
  \bibfield  {author} {\bibinfo {author} {\bibfnamefont {Rudnei}\ \bibnamefont {O.~Ramos}}\ and\ \bibinfo {author} {\bibfnamefont {Gabriel~S.}\ \bibnamefont {Rodrigues}},\ }\bibfield  {title} {\enquote {\bibinfo {title} {{Viability of warm inflation with standard model interactions}},}\ }\href {\doibase 10.1103/wn1m-19gt} {\bibfield  {journal} {\bibinfo  {journal} {Phys. Rev. D}\ }\textbf {\bibinfo {volume} {111}},\ \bibinfo {pages} {123527} (\bibinfo {year} {2025})},\ \Eprint {http://arxiv.org/abs/2504.20943} {arXiv:2504.20943 [hep-ph]} \BibitemShut {NoStop}%
\bibitem [{\citenamefont {Laine}\ \emph {et~al.}(2025)\citenamefont {Laine}, \citenamefont {Procacci},\ and\ \citenamefont {Rogelj}}]{Laine:2025rll}%
  \BibitemOpen
  \bibfield  {author} {\bibinfo {author} {\bibfnamefont {M.}~\bibnamefont {Laine}}, \bibinfo {author} {\bibfnamefont {S.}~\bibnamefont {Procacci}}, \ and\ \bibinfo {author} {\bibfnamefont {A.}~\bibnamefont {Rogelj}},\ }\bibfield  {title} {\enquote {\bibinfo {title} {{Evolution of coupled scalar perturbations through smooth reheating. II. Thermal fluctuation regime}},}\ }\href@noop {} {\  (\bibinfo {year} {2025})},\ \Eprint {http://arxiv.org/abs/2507.12849} {arXiv:2507.12849 [hep-ph]} \BibitemShut {NoStop}%
\bibitem [{\citenamefont {Graham}\ and\ \citenamefont {Moss}(2009)}]{Graham:2009bf}%
  \BibitemOpen
  \bibfield  {author} {\bibinfo {author} {\bibfnamefont {Chris}\ \bibnamefont {Graham}}\ and\ \bibinfo {author} {\bibfnamefont {Ian~G.}\ \bibnamefont {Moss}},\ }\bibfield  {title} {\enquote {\bibinfo {title} {{Density fluctuations from warm inflation}},}\ }\href {\doibase 10.1088/1475-7516/2009/07/013} {\bibfield  {journal} {\bibinfo  {journal} {JCAP}\ }\textbf {\bibinfo {volume} {07}},\ \bibinfo {pages} {013} (\bibinfo {year} {2009})},\ \Eprint {http://arxiv.org/abs/0905.3500} {arXiv:0905.3500 [astro-ph.CO]} \BibitemShut {NoStop}%
\bibitem [{\citenamefont {Bastero-Gil}\ \emph {et~al.}(2011)\citenamefont {Bastero-Gil}, \citenamefont {Berera},\ and\ \citenamefont {Ramos}}]{Bastero-Gil:2011rva}%
  \BibitemOpen
  \bibfield  {author} {\bibinfo {author} {\bibfnamefont {Mar}\ \bibnamefont {Bastero-Gil}}, \bibinfo {author} {\bibfnamefont {Arjun}\ \bibnamefont {Berera}}, \ and\ \bibinfo {author} {\bibfnamefont {Rudnei~O.}\ \bibnamefont {Ramos}},\ }\bibfield  {title} {\enquote {\bibinfo {title} {{Shear viscous effects on the primordial power spectrum from warm inflation}},}\ }\href {\doibase 10.1088/1475-7516/2011/07/030} {\bibfield  {journal} {\bibinfo  {journal} {JCAP}\ }\textbf {\bibinfo {volume} {07}},\ \bibinfo {pages} {030} (\bibinfo {year} {2011})},\ \Eprint {http://arxiv.org/abs/1106.0701} {arXiv:1106.0701 [astro-ph.CO]} \BibitemShut {NoStop}%
\bibitem [{\citenamefont {Montefalcone}\ \emph {et~al.}(2024)\citenamefont {Montefalcone}, \citenamefont {Aragam}, \citenamefont {Visinelli},\ and\ \citenamefont {Freese}}]{Montefalcone:2023pvh}%
  \BibitemOpen
  \bibfield  {author} {\bibinfo {author} {\bibfnamefont {Gabriele}\ \bibnamefont {Montefalcone}}, \bibinfo {author} {\bibfnamefont {Vikas}\ \bibnamefont {Aragam}}, \bibinfo {author} {\bibfnamefont {Luca}\ \bibnamefont {Visinelli}}, \ and\ \bibinfo {author} {\bibfnamefont {Katherine}\ \bibnamefont {Freese}},\ }\bibfield  {title} {\enquote {\bibinfo {title} {{WarmSPy: a numerical study of cosmological perturbations in warm inflation}},}\ }\href {\doibase 10.1088/1475-7516/2024/01/032} {\bibfield  {journal} {\bibinfo  {journal} {JCAP}\ }\textbf {\bibinfo {volume} {01}},\ \bibinfo {pages} {032} (\bibinfo {year} {2024})},\ \Eprint {http://arxiv.org/abs/2306.16190} {arXiv:2306.16190 [astro-ph.CO]} \BibitemShut {NoStop}%
\bibitem [{\citenamefont {Laine}\ \emph {et~al.}(2024)\citenamefont {Laine}, \citenamefont {Procacci},\ and\ \citenamefont {Rogelj}}]{Laine:2024wyv}%
  \BibitemOpen
  \bibfield  {author} {\bibinfo {author} {\bibfnamefont {M.}~\bibnamefont {Laine}}, \bibinfo {author} {\bibfnamefont {S.}~\bibnamefont {Procacci}}, \ and\ \bibinfo {author} {\bibfnamefont {A.}~\bibnamefont {Rogelj}},\ }\bibfield  {title} {\enquote {\bibinfo {title} {{Evolution of coupled scalar perturbations through smooth reheating. Part I. Dissipative regime}},}\ }\href {\doibase 10.1088/1475-7516/2024/10/040} {\bibfield  {journal} {\bibinfo  {journal} {JCAP}\ }\textbf {\bibinfo {volume} {10}},\ \bibinfo {pages} {040} (\bibinfo {year} {2024})},\ \Eprint {http://arxiv.org/abs/2407.17074} {arXiv:2407.17074 [hep-ph]} \BibitemShut {NoStop}%
\bibitem [{\citenamefont {Rodrigues}\ and\ \citenamefont {O.~Ramos}(2025)}]{Rodrigues:2025neh}%
  \BibitemOpen
  \bibfield  {author} {\bibinfo {author} {\bibfnamefont {Gabriel~S.}\ \bibnamefont {Rodrigues}}\ and\ \bibinfo {author} {\bibfnamefont {Rudnei}\ \bibnamefont {O.~Ramos}},\ }\bibfield  {title} {\enquote {\bibinfo {title} {{WI2easy: warm inflation dynamics made easy}},}\ }\href@noop {} {\  (\bibinfo {year} {2025})},\ \Eprint {http://arxiv.org/abs/2504.17760} {arXiv:2504.17760 [astro-ph.CO]} \BibitemShut {NoStop}%
\bibitem [{\citenamefont {Rubio}\ and\ \citenamefont {Tomberg}(2019)}]{Rubio:2019ypq}%
  \BibitemOpen
  \bibfield  {author} {\bibinfo {author} {\bibfnamefont {Javier}\ \bibnamefont {Rubio}}\ and\ \bibinfo {author} {\bibfnamefont {Eemeli~S.}\ \bibnamefont {Tomberg}},\ }\bibfield  {title} {\enquote {\bibinfo {title} {{Preheating in Palatini Higgs inflation}},}\ }\href {\doibase 10.1088/1475-7516/2019/04/021} {\bibfield  {journal} {\bibinfo  {journal} {JCAP}\ }\textbf {\bibinfo {volume} {04}},\ \bibinfo {pages} {021} (\bibinfo {year} {2019})},\ \Eprint {http://arxiv.org/abs/1902.10148} {arXiv:1902.10148 [hep-ph]} \BibitemShut {NoStop}%
\bibitem [{\citenamefont {Husdal}(2016)}]{Husdal:2016haj}%
  \BibitemOpen
  \bibfield  {author} {\bibinfo {author} {\bibfnamefont {Lars}\ \bibnamefont {Husdal}},\ }\bibfield  {title} {\enquote {\bibinfo {title} {{On Effective Degrees of Freedom in the Early Universe}},}\ }\href {\doibase 10.3390/galaxies4040078} {\bibfield  {journal} {\bibinfo  {journal} {Galaxies}\ }\textbf {\bibinfo {volume} {4}},\ \bibinfo {pages} {78} (\bibinfo {year} {2016})},\ \Eprint {http://arxiv.org/abs/1609.04979} {arXiv:1609.04979 [astro-ph.CO]} \BibitemShut {NoStop}%
\end{thebibliography}%

\end{document}